\documentclass[submitting]{nst}

\usepackage{amssymb}
\usepackage{graphicx}
\usepackage{graphics}
\usepackage{amsmath}
\usepackage{placeins}
\usepackage{hyperref}
\usepackage{multirow}
\usepackage{xcolor}

\usepackage{subfigure,dcolumn}
\usepackage{epstopdf}
\usepackage{mhchem}
\usepackage{upgreek}
\usepackage[T2A,T1]{fontenc}
\usepackage{listings}
\begin{document}

\title{Development of a quadripartite wakefield structure as dechirper for free electron laser}

\author{Yu Ji}
\affiliation{Institute of Advanced Light Source Facilities, Shenzhen, China 518107}

\author{Congrui Lei}
\affiliation{Institute of Advanced Light Source Facilities, Shenzhen, China 518107}

\author{Jiahang Shao}
\email[Corresponding author, ]{shaojiahang@mail.iasf.ac.cn}
\affiliation{Institute of Advanced Light Source Facilities, Shenzhen, China 518107}

\author{Yong Yu}
\affiliation{Institute of Advanced Light Source Facilities, Shenzhen, China 518107}

\author{Jitao Sun}
\affiliation{Institute of Advanced Light Source Facilities, Shenzhen, China 518107}

\author{Xiaofan Wang}
\affiliation{Institute of Advanced Light Source Facilities, Shenzhen, China 518107}

\author{Huaiqian Yi}
\affiliation{Institute of Advanced Light Source Facilities, Shenzhen, China 518107}

\author{Zongbin Li}
\affiliation{Institute of Advanced Light Source Facilities, Shenzhen, China 518107}

\author{Lei He}
\affiliation{Institute of Advanced Light Source Facilities, Shenzhen, China 518107}

\author{Hongfei Wang}
\affiliation{Institute of Advanced Light Source Facilities, Shenzhen, China 518107}

\author{Jianping Wei}
\affiliation{Institute of Advanced Light Source Facilities, Shenzhen, China 518107}

\author{Wei Wei}
\affiliation{Institute of Advanced Light Source Facilities, Shenzhen, China 518107}

\author{Wei Wang}
\affiliation{Institute of Advanced Light Source Facilities, Shenzhen, China 518107}

\author{Jiayue Yang}
\affiliation{Dalian Institute of Chemical Physics, Chinese Academy of Sciences, Dalian, China 116023}

\author{Weiqing Zhang}
\affiliation{Dalian Institute of Chemical Physics, Chinese Academy of Sciences, Dalian, China 116023}

\author{Xueming Yang}
\affiliation{Dalian Institute of Chemical Physics, Chinese Academy of Sciences, Dalian, China 116023}

\begin{abstract}
	Wakefield structures are critical for beam manipulation in free-electron lasers (FELs), particularly when serving as dechirpers, where beam-induced longitudinal wakefields compensate the energy chirp introduced during beam magnetic compression. However, conventional planar structures also generate time-dependent quadrupole wakefields due to their asymmetric geometry, which can cause beam mismatch and projected emittance growth. To address this limitation, we propose a quadripartite wakefield structure comprising four identical corrugated plates, able to fully suppress quadrupole wakefields while preserving strong longitudinal wakefields. To accurately evaluate its performance, we calculate wake potentials based on the Panofsky-Wenzel theorem using three-dimensional simulation software and extract the corresponding wake functions by deconvolution. We further adopt a particle-to-particle (P2P) tracking method incorporating these wake functions, which is capable of accounting for higher-order components and nonlinear effects that are typically neglected in standard tracking codes. Simulation results confirm that the quadripartite geometry offers significantly reduced projected emittance growth and a 25\% shorter structure length compared with the planar design. The tracking method also reveals that the nonlinearities of three-dimensional wakefields induce noticeable slice emittance growth for large transverse beam sizes, which may in turn affect lasing performance. In addition, the tracking method enables analysis of various types of assembly error and indicates that misalignment along the direction of plate motion may severely degrade the emittance via dipole wakefields. Such misalignment can be mitigated through beam-based alignment and precise plate adjustment using high-resolution servo motors.
\end{abstract}


\maketitle

\nolinenumbers

\section{Introduction}\label{sec.I}
Wakefields are induced when a charged bunch traverses a corrugated or dielectric structure. Structure-based wakefield acceleration represents a promising approach to achieve gradients significantly higher than those attained by conventional techniques~\cite{OShea_NC2016,Jing_NIMA2018,Shao_PRAB2020}. Furthermore, wakefields have been demonstrated to be effective tools for beam manipulation and diagnostics in FELs, where the lasing performance strongly depends on beam properties~\cite{Feng_NST2018}. Short-range wakefields from the bunch head alter the longitudinal or transverse momentum of the tail, tailoring the beam for various applications. Typically, wakefield structures are employed as dechirpers to compensate the linear energy chirp introduced during magnetic bunch compression~\cite{Bane_NIMA2012,Emma_PRL2014,Zhang_PRAB2015,Gong_NST2021}. These structures have also been adapted for a broader range of beam manipulation and diagnostics, such as passive linearization~\cite{Craievich_PRAB2010,Deng_PRL2014,Penco_PRL2017,Wang_NST2018,Ding_Note2018,Mayet_PRAB2020}, lasing slice and spectrum control~\cite{Lutman_NP2016,Bettoni_PRAB2016B,Zagorodnov_NIM2016,Qin_PRAB2017,Chao_PRL2018,Lutman_PRL2018,Bettoni_PRAB2021,Duris_PRL2021}, passive deflection~\cite{Bettoni_PRAB2016A,Seok_PRAB2018,Dijkstal_PRR2022,Qin_PRAB2023,Dijkstal_PRAB2024}, and THz generation~\cite{Stupakov_PRAB2015,Floettmann_JSR2021}. 

In addition to structure-based wakefield devices, plasma-based dechirpers have recently been applied to reduce the energy spread of plasma accelerator to the level of 0.1\% (full width at half maximum), which could have a strong impact on future FEL research and development~\cite{Wu_PRL2019,Liu_PRL2024}.

The Shenzhen Superconducting Soft X-ray Free-Electron Laser (S$^{3}$FEL) is a newly proposed high-repetition-rate FEL facility featuring multiple undulator lines that lase in the 1-30~nm range~\cite{Wang_IPAC2023}. Civil construction for the FEL infrastructure has already started and key technologies are under intense development~\cite{Sun_PRR2024,Huang_IBIC2024,Zhu_JI2025,Hu_CEC2021,Zhang_FP2023,Xu_SR2023,Wu_JSR2024}. Wakefield structures are investigated to serve as dechirpers and as fundamental components for advanced lasing modes. 

Wakefield structures with various geometries, including round~\cite{Bane_NIMA2012}, planar~\cite{Zhang_PRAB2015,Lutman_PRL2018}, L-shaped~\cite{Qin_PRAB2023}, and square~\cite{Ganter_FEL2017}, have been proposed or applied in FEL facilities. The planar structure with two movable corrugated plates is prevalent among these configurations since the wakefield strength can be adjusted by varying the gap, which offers tuning flexibility and enables novel applications. However, quadrupole wakefield is generated in addition to longitudinal wakefield, and such time-dependent transverse wakefield can cause beam mismatch and projected emittance growth. Therefore, it usually requires quadrupole wakefield compensation by placing two planar structures orthogonally and maintaining nearly identical average $\beta$ functions throughout the structures~\cite{Zhang_PRAB2015}, which introduces complication in beam tuning and optimization.

In planar structures, the quadrupole wakefield is caused by quadrupole components of monopole modes  due to the asymmetric geometry. We therefore propose a novel quadripartite wakefield structure consisting of four identical corrugated plates~\cite{Ji_LINAC2024}. The wakefield can be flexibly adjusted by tuning the gaps between orthogonal pairs, similar to planar structures. Meanwhile, the quadrupole wakefield is naturally canceled with identical horizontal and vertical gaps owing to the symmetric geometry, leading to preserved emittance. These properties make the quadripartite design a promising candidate for S$^3$FEL wakefield structures.

Accurate short-range wakefields are critical for reliably assessing the performance of the quadripartite structure. For simple geometries, such as round or planar plates, wakefields can be analytically calculated by field matching techniques~\cite{Ng_PRD1990,Xiao_PRE2001,Bane_PRAB2003,Zhang_PRAB2015} or surface impedance models~\cite{Bane_PRAB2015,Bane_NIMA2016,Bane_PRAB2016}, or numerically simulated by two-dimensional codes~\cite{Zagorodnov_PRAB2015,Novokhatski_PRAB2015}. The conformal mapping method~\cite{Baturin_PRAB2016,Qin_PRAB2023} is also applicable to waveguides with arbitrary cross sections, provided they are enclosed by boundaries with uniform surface impedance. However, due to the complex boundary conditions of the proposed quadripartite structure, analytical approaches are not feasible, while full three-dimensional numerical simulations~\cite{Bruns_IEEE1996,Ng_NAPAC2013,Zagorodnov_NAPAC2016,CST} are required to obtain the wakefields. 

In addition, beam dynamics simulations based on the resolved wakefields are essential for estimating the structure's performance in FEL applications. Conventional tracking codes~\cite{Elegant,ASTRA} typically rely on one-dimensional wakefields near the beam axis and are not capable of accounting for higher-order components and nonlinear effects that are revealed by three-dimensional simulations. To address this limitation, we adopt a particle-to-particle tracking approach~\cite{Wang_Thesis} incorporating three-dimensional wake functions deconvoluted from numerical simulations. This method enables the study of nonlinear wakefield effects within the structure and allows for performance analysis under various types of assembly error.

For the intended application as dechirpers in S$^{3}$FEL, three-dimensional wakefield simulations and particle-to-particle tracking demonstrate that the quadripartite geometry provides significantly reduced projected emittance growth and a shorter structure length compared with the conventional planar design. Although their impact on beam dynamics and lasing behavior is negligible under the nominal beam conditions, nonlinearities of the three-dimensional wakefields induce slice emittance growth with larger transverse beam sizes which can degrade the lasing performance. The results further indicate that the impact of assembly errors on beam dynamics can be effectively mitigated, confirming the technical feasibility of the quadripartite structure.

This manuscript is organized as follows. Section~\ref{sec.II} describes the geometry and parameters of both the planar and quadripartite wakefield structures. Section~\ref{sec.III} outlines the methodology for calculating one-dimensional and three-dimensional wakefields using \texttt{ECHO3D}~\cite{Zagorodnov_NAPAC2016}. The wakefield properties and performance for dechirper application are compared between the two structures based on one-dimensional wakefield using \texttt{ELEGANT}~\cite{Elegant} in Sec.~\ref{sec.IV}. Section\ref{sec.V} presents the three-dimensional wakefields and emphasizes the difference from the one-dimensional ones. Section~\ref{sec.VI} details the particle-to-particle model and corresponding beam dynamics and lasing simulations. Assembly error analysis and mitigation strategies are discussed in Sec.~\ref{sec.VIII}. Section~\ref{sec.IX} introduces a preliminary mechanical design. Finally, Section~\ref{sec.XI} concludes the study with a summary of the key findings and directions for future work.

\section{Geometries of Wakefield Structures}\label{sec.II}
\begin{figure}[!htb]
	\centering
	\includegraphics*[width=1\columnwidth]{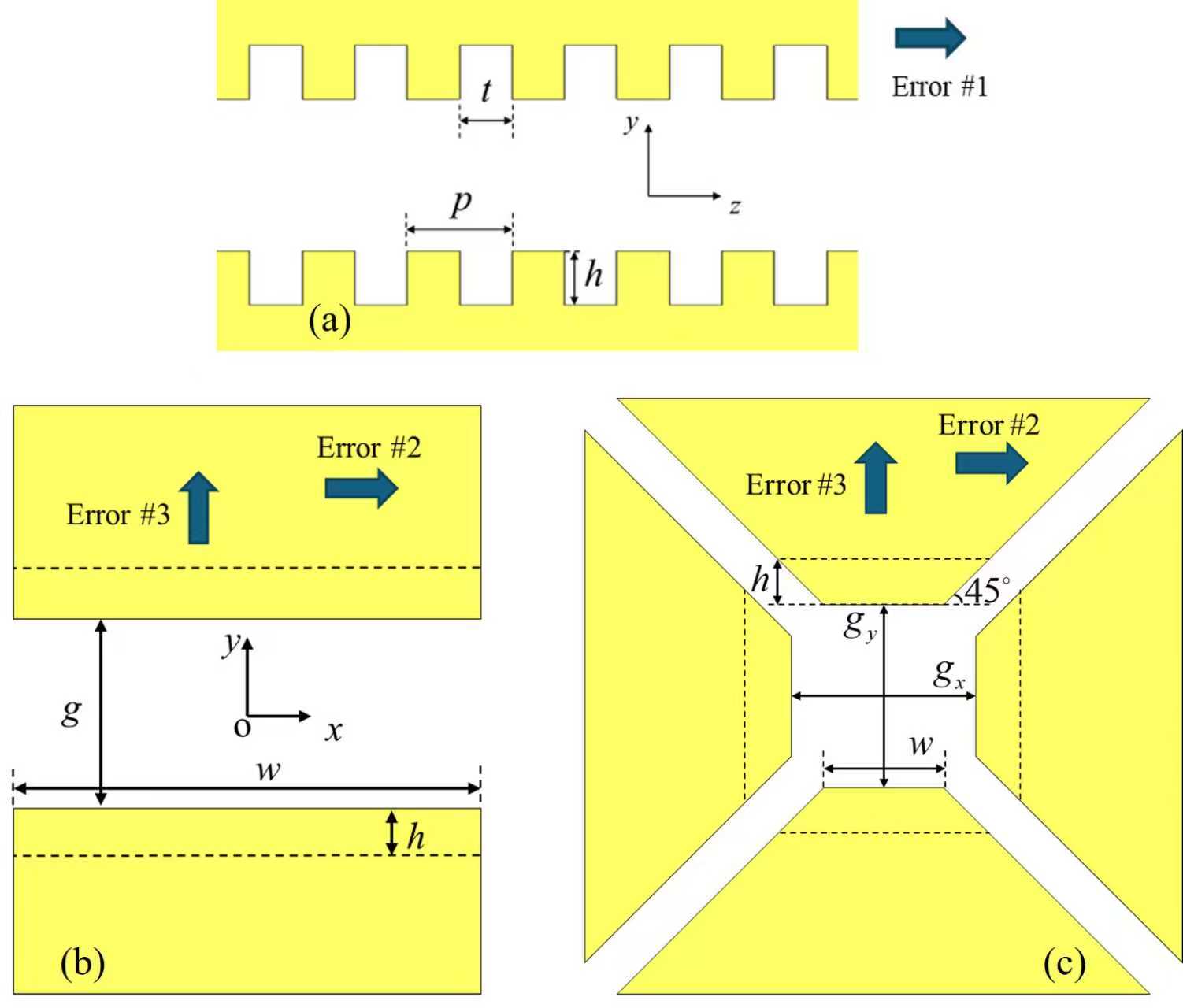}
	\caption{Cross sections of the wakefield structures, in which the types of shift assembly error to be discussed in Sec.~\ref{sec.VIII} are marked. (a) Side view. (b) Front view of the planar structure. (c) Front view of the quadripartite structure.}
	\label{fig_crosssection}
\end{figure}

The cross sections of both the planar structure and the proposed quadripartite structure are illustrated in Fig.~\ref{fig_crosssection}. For both structures, the corrugation width is defined as $w$. The effective gap (beam aperture) between the opposite plates is defined as $g$ in the planar structure, and $g_{x}$ and $g_{y}$ for the corresponding directions in the quadripartite structure. It should be noted that the minimal effective gaps in the quadripartite structure are limited by the gaps between neighboring corrugation plates that should be larger than 100~${\mu}$m for  mechanical and vacuum considerations. 

In this study, the corrugation parameters and the nominal effective gap are adopted from the dechirpers applied in LCLS~\cite{Zhang_PRAB2015}. These values in the quadripartite structure are set to be identical to those in the planar one for direct comparison, as listed in Table~\ref{tab_parameters}.

\begin{table}[!htb]
    \centering
    \caption{Geometry parameters of the wakefield structures.}
    \label{tab_parameters}
    \begin{ruledtabular}
	    \begin{tabular}{c c c}
	        Parameter&\multicolumn{2}{c}{Value (mm)}\\
	        \hline
	        Corrugation period \(p\) &\multicolumn{2}{c}{0.5}\\
	        Corrugation length \(t\) &\multicolumn{2}{c}{0.25}\\
	        Corrugation depth \(h\) &\multicolumn{2}{c}{0.5}\\
	        & Planar & Quadripartite \\ 
	        Corrugation width \(w\) & 12 & 1\\
	        Nominal effective gap & 1.4 & 1.4 (when $g_{x}=g_{y}$)\\
	        Minimal effective gap & 0.1 & 1.2 (when $g_{x}=g_{y}$)\\   
	    \end{tabular}
	\end{ruledtabular}
\end{table}

\section{Calculation of wakefield by numerical simulation}\label{sec.III}
The simulation in this section adopts the planar structure with the nominal effective gap and an on-axis beam with Gaussian temporal distribution. The root-mean-square (RMS) bunch length $\sigma$ is set to 15~$\mu$m, similar to the realistic beam at the dechirper location in S$^3$FEL.

For a line-charge bunch with an off-axis position ($x_{0}$, $y_{0}$) in a three-dimensional structure, the code \texttt{ECHO3D} is capable of computing the longitudinal wake potential $W_{\|}$ at arbitrary transverse position ($x$, $y$) in the time domain. More information on the \texttt{ECHO3D} simulations and their validation against the commercial three-dimensional software \texttt{CST}~\cite{CST} is provided in Appendix~\ref{ECHO}.

\subsection{Three-dimensional wakefield}\label{sec.III.A}
Provided longitudinal wake potentials, transverse wake potentials can be then determined according to the Panofsky-Wenzel theorem~\cite{PW_RSI1956} as follows:
\begin{widetext}
\begin{equation}
	\label{eq_PW}
	\left\{
	\begin{aligned}
	W_{x}(x_{0}, x, y_{0}, y, s)&=\frac{\partial}{\partial x} \int W_{\|}(x_{0}, x, y_{0}, y, s)ds=\lim _{\Delta x \rightarrow 0} \int \frac{W_{\|}(x_{0}, x+\Delta x, y_{0}, y, s)-W_{\|}(x_{0}, x, y_{0}, y, s)}{\Delta x}ds\\
	W_{y}(x_{0}, x, y_{0}, y, s)&=\frac{\partial}{\partial y} \int W_{\|}(x_{0}, x, y_{0}, y, s)ds=\lim _{\Delta y \rightarrow 0} \int \frac{W_{\|}(x_{0}, x, y_{0}, y+\Delta y, s)-W_{\|}(x_{0}, x, y_{0}, y, s)}{\Delta y}ds
	\end{aligned}
	\right.,
\end{equation}
\end{widetext}
where $s$ represents the longitudinal position behind the head of the bunch. $W_{\|}$, $W_{x}$, and $W_{y}$ are normalized to charge and structure length with unit in $V/(pc \times m)$.

To ensure the accuracy of numerical results, it is crucial to identify appropriate mesh sizes $(dx, dy, dz)$ in the \texttt{ECHO3D} simulation as well as $(\Delta x, \Delta y)$ in Eq.\eqref{eq_PW}.  $(\Delta x, \Delta y)$ should be taken as integer multiples of $(dx, dy)$.

\begin{figure}[!htb]
	\centering
	\includegraphics[width=\hsize]{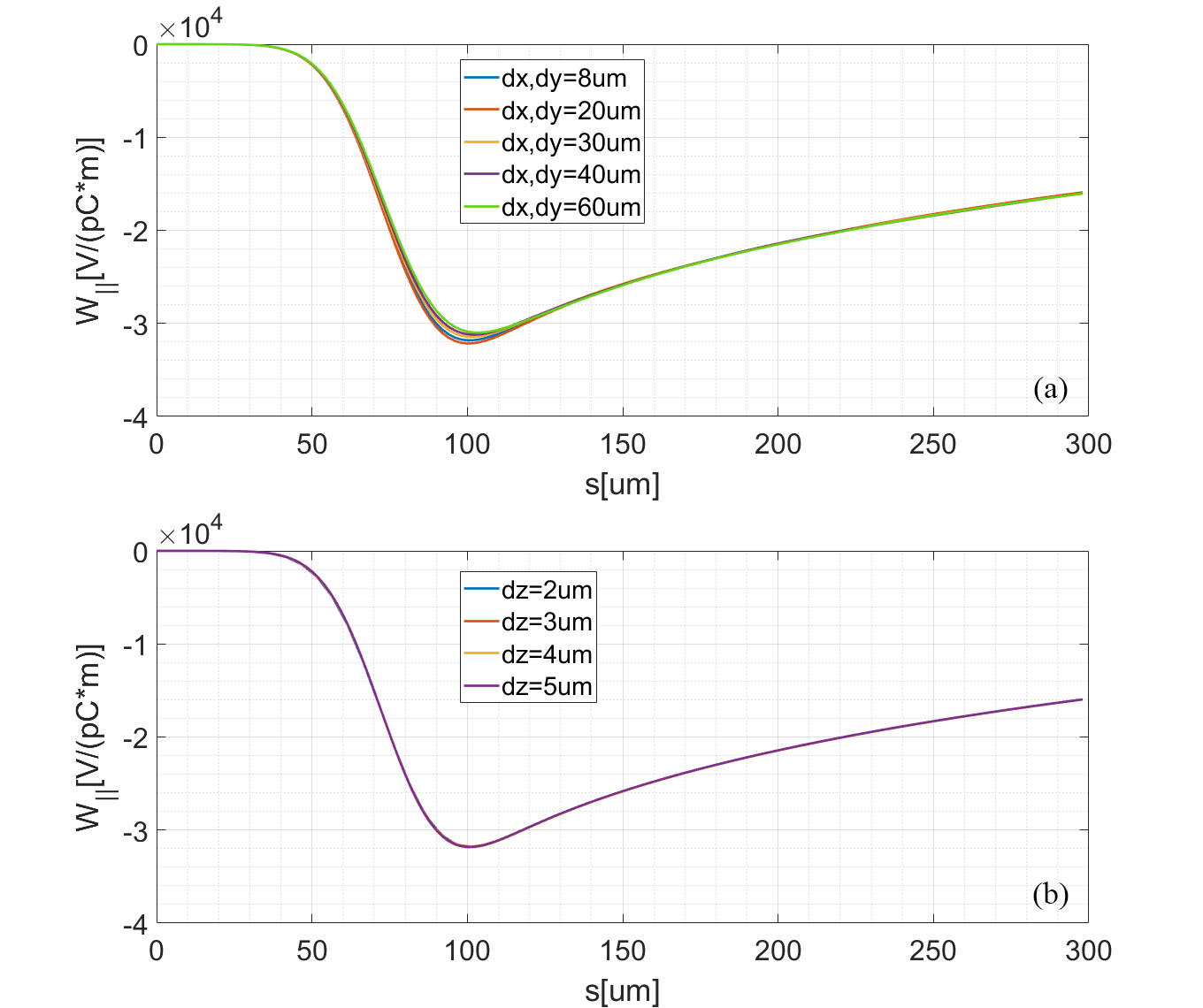}
	\caption{Convergence of the longitudinal wake potential with mesh sizes in \texttt{ECHO3D} simulations. (a) Fix $dz=2~\mu$m. (b) Fix $dx=dy=8~\mu$m. }
	\label{fig_converge_longitudinal}
\end{figure}

A calculation example of the longitudinal wake potential is illustrated in Fig.~\ref{fig_converge_longitudinal}, where $W_{\|}(0, 0, 0, 0, s)$ is calculated with various mesh sizes. With the recommended $dz$ by \texttt{ECHO3D} (<1/5 of RMS bunch length), the longitudinal wake potential slightly varies with the transverse mesh size. Meanwhile, the longitudinal wake potential nearly stays constant when changing $dz$ by fixing $dx=dy=8$~$\mu$m.

\begin{figure}[!htb]
	\centering
	\includegraphics[width=\hsize]{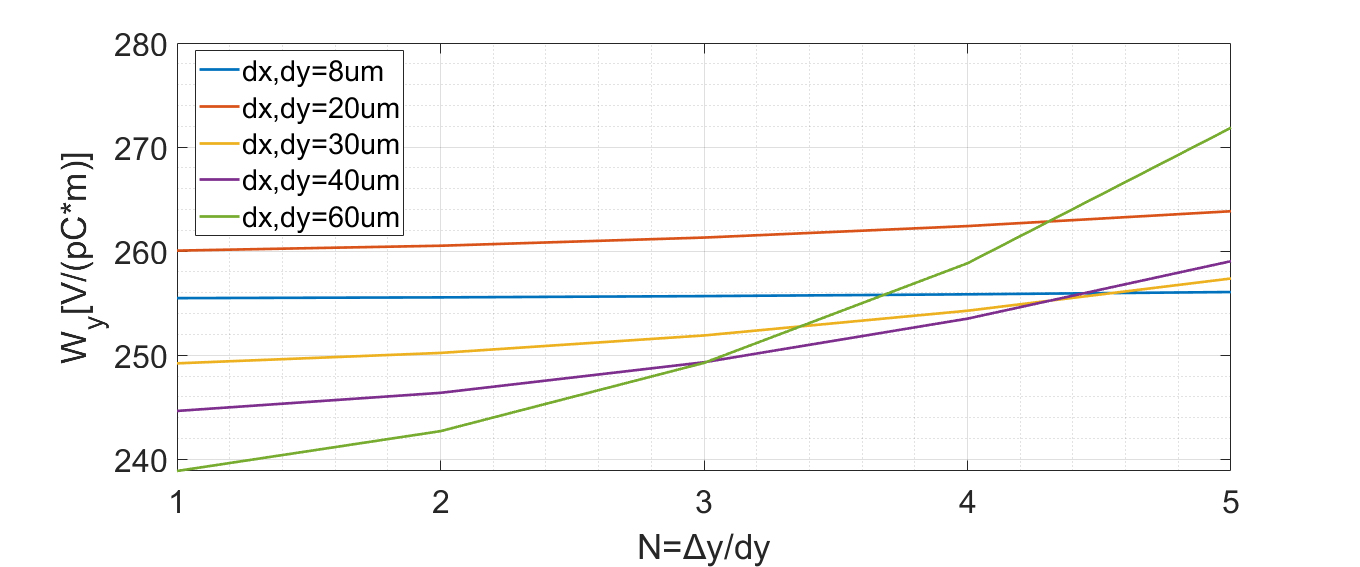}
	\caption{Convergence of the transverse wake potential with various simulation and calculation conditions.}
	\label{fig_converge_transverse}
\end{figure}

\begin{figure}[!htb]
	\centering
	\includegraphics[width=\hsize]{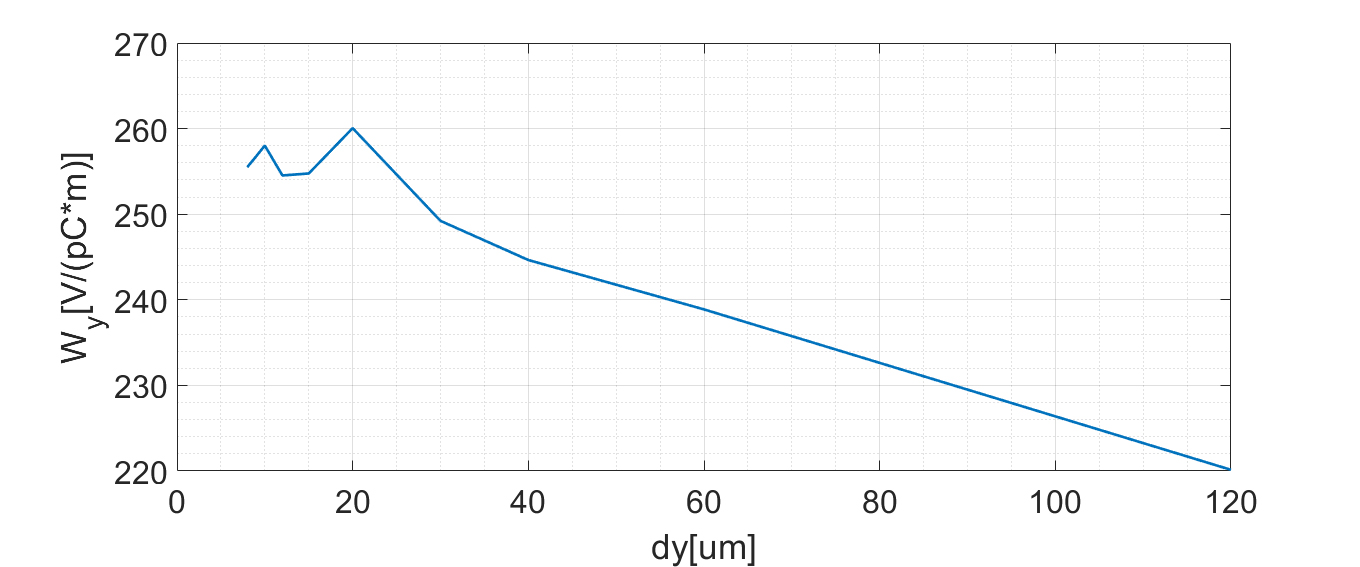}
	\caption{Convergence of the transverse wake potential when $\Delta y = dy$.}
	\label{fig_converge_transverse_N}
\end{figure}

Another example of the transverse wake potential is illustrated in Fig.~\ref{fig_converge_transverse} and Fig.~\ref{fig_converge_transverse_N}, where $W_{y}(0, 0, 0, 120~\mu m, 100~\mu m)$ is calculated with various conditions. With fixed mesh sizes, $W_{y}$ converges when $\Delta y$ approaches $dy$ (Fig.~\ref{fig_converge_transverse}). Meanwhile, $W_{y}$ converges with small mesh sizes when $\Delta y$ is set to $dy$ (Fig.~\ref{fig_converge_transverse_N}). The oscillation observed when $dy$ is smaller than 20~$\mu$m may originate from numerical noise.

Therefore, ($dx, dy, dz$) is set to (8~$\mu$m, 8~$\mu$m, 2~$\mu$m) and ($\Delta x, \Delta y$) is set to (8~$\mu$m, 8~$\mu$m) in the following simulations.

\subsection{One-dimensional wakefield}
Conventional beam dynamics codes apply one-dimensional wakefields during tracking. The one-dimensional longitudinal wakefield $W_{\|}(s)$ uses the on-axis results $W_{\|}(0, 0, 0, 0, s)$ and is assumed to be independent of ($x_0, y_0, x, y$). Meanwhile, the one-dimensional transverse wakefields are simplified as constant dipole and quadrupole components which are linear with ($x_0, y_0$) and ($x, y$), respectively.

The dipole and quadrupole components can be derived from near-axis wakefield by Taylor expansion under the first-order approximation as follows:
\begin{widetext}
	\begin{equation}
		\label{eq_Taylor}
		\left\{
		\begin{aligned}
			W_{x}\left(x_{0}, x, 0, 0, s\right)=&W_{x}\left(0, 0, 0, 0, s\right)
			+x_{0} \frac{\partial}{\partial x_{0}} W_{x}\left(x_0, x, 0, 0, s\right)|_{x_0=0, x=0}
			+x \frac{\partial}{\partial x} W_{x}\left(x_0, x, 0, 0, s\right)|_{x_0=0, x=0}+O(n)\\
			\equiv&W_{x}\left(0, 0, 0, 0, s\right)+W_{dx}(s)x_{0} +W_{qx}(s)x +O(n)\\
			W_{y}\left(0, 0, y_{0}, y, s\right)=&W_{y}\left(0, 0, 0, 0, s\right)
			+y_{0} \frac{\partial}{\partial y_{0}} W_{y}\left(0, 0, y_{0}, y, s\right)|_{y_0=0, y=0}
			+y \frac{\partial}{\partial y} W_{y}\left(0, 0, y_{0}, y, s\right)|_{y_0=0, y=0}+O(n)\\
			\equiv&W_{y}\left(0, 0, 0, 0, s\right)+W_{dy}(s)y_{0} +W_{qy}(s)y +O(n)
		\end{aligned}
		\right.,
	\end{equation}
\end{widetext}
where $W_{d}$ and $W_{q}$ denote the dipole and the quadrupole wakefield, respectively. $W_{d}$ and $W_{q}$ are normalized to charge, structure length, and beam offset with unit in $V/(pc \times mm \times m)$.

Therefore, the dipole and quadrupole wakefields can be derived as
\begin{equation}
	\label{eq_1D}
	\left\{
	\begin{aligned}
		W_{dx}(s)&=\lim _{\Delta x_{0} \rightarrow 0} \frac{W_{x}(\Delta x_{0}, 0, 0, 0, s)-W_{x}(0, 0, 0, 0, s)}{\Delta x_{0}}\\
		W_{qx}(s)&=\lim _{\Delta x \rightarrow 0} \frac{W_{x}(0, \Delta x, 0, 0, s)-W_{x}(0, 0, 0, 0, s)}{\Delta x} \\
		W_{dy}(s)&=\lim _{\Delta y_{0} \rightarrow 0} \frac{W_{y}(0, 0, \Delta y_{0}, 0, s)-W_{y}(0, 0, 0, 0, s)}{\Delta y_{0}}\\
		W_{qy}(s)&=\lim _{\Delta y \rightarrow 0} \frac{W_{y}(0, 0, 0, \Delta y, s)-W_{y}(0, 0, 0, 0, s)}{\Delta y} \\
	\end{aligned}
	\right.,
\end{equation}
where \(W_{dx}=W_{dy}\) and \(W_{qx}=-W_{qy}\).

Following the same convergence study procedure, $\Delta x_{0}, \Delta y_{0}, \Delta x, \Delta y$ in Eq.~\eqref{eq_1D} are all set to 8~$\mu$m in following analysis.

\subsection{Wake function calculation}
Wake functions, rather than wake potentials, are essential for beam dynamics study with arbitrary temporal profiles. Since \texttt{ECHO3D} currently can only simulate wakefields potentials for line-charge beam with Gaussian temporal distributions $G(z)$, it is necessary to extract wake functions by deconvolution according to
\begin{equation}
	\label{eq_deconv}
	\left\{
	\begin{aligned}	
		W_{\|}(x_{0}, x, y_{0}, y, s)=&\int_{0}^{\infty} w_{\|}(x_{0}, x, y_{0}, y, z) G(s-z) d z \\
		W_{x}(x_{0}, x, y_{0}, y, s)=&\int_{0}^{\infty} w_{x}(x_{0}, x, y_{0}, y, z) G(s-z) d z \\
		W_{y}(x_{0}, x, y_{0}, y, s)=&\int_{0}^{\infty} w_{y}(x_{0}, x, y_{0}, y, z) G(s-z) d z \\
	\end{aligned}
	\right.,
\end{equation}
where ($w_{\|}, w_x, w_y$) denote the three-dimensional wake functions. The one-dimensional wake functions can be therefore calculated following the same derivation in Eq.~\eqref{eq_1D}.

\begin{figure}[!htb]
	\centering
	\includegraphics[width=\hsize]{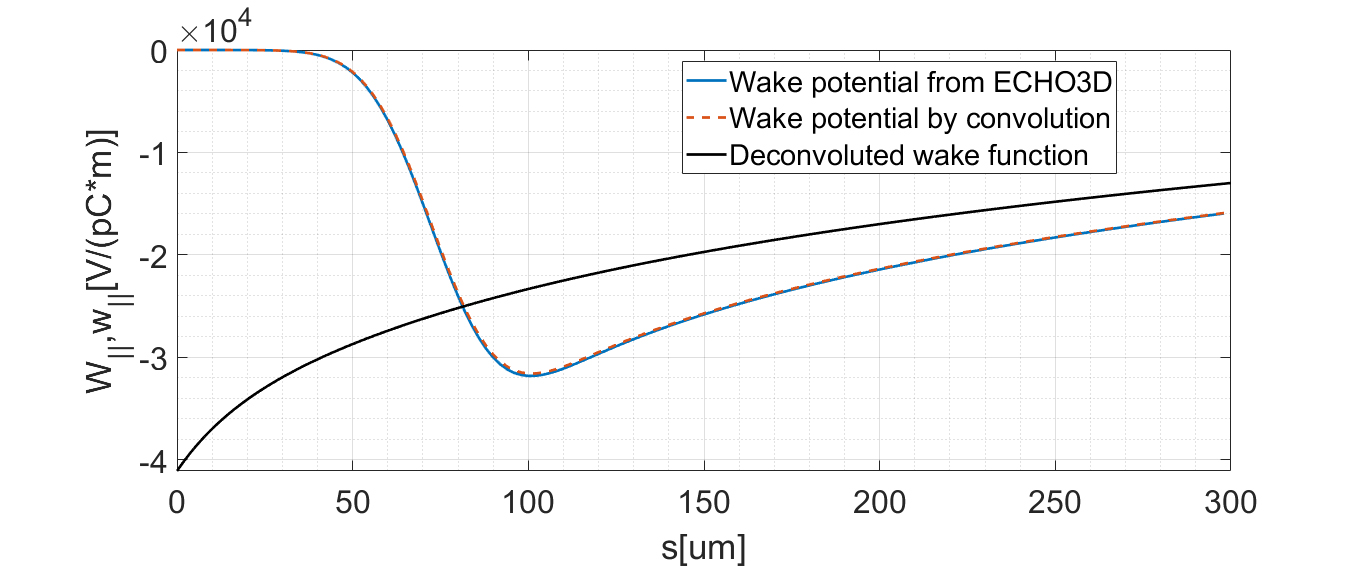}
	\caption{Validation of the deconvolution method.}
	\label{fig_deconv}
\end{figure}

Validation of the deconvolution is illustrated in Fig.~\ref{fig_deconv}, where the original wake potential from \texttt{ECHO3D} simulation (blue solid line) and reconstructed wake potential from the deconvoluted wake function and the Gaussian temporal distribution (red dashed line) overlaps with each other. This validation also extends to transverse wakefields according to the integral and differential properties of convolution. 

It should be noted that deconvoluted wake functions may vary with the applied deconvolution methods, especially the head part within $1 \times \sigma$ range. In fact, short-range wake functions have been approximated as different forms in previous studies~\cite{Zhang_PRAB2015,Bane_NIMA2016,Bane_PRAB2016,Qin_PRAB2023}. Such differences may impact beam dynamics of a single particle, whereas the influence on the entire beam is negligible in terms of statistic analysis and would not affect the following analysis.

\section{Structure comparison with one-dimensional wakefields}\label{sec.IV}
Based on the methodology in the previous section, the two structures are compared with one-dimensional wake functions in this section.

\subsection{Wakefield calculation results}
\begin{figure}[!htb]
	\centering
	\includegraphics[width=\hsize]{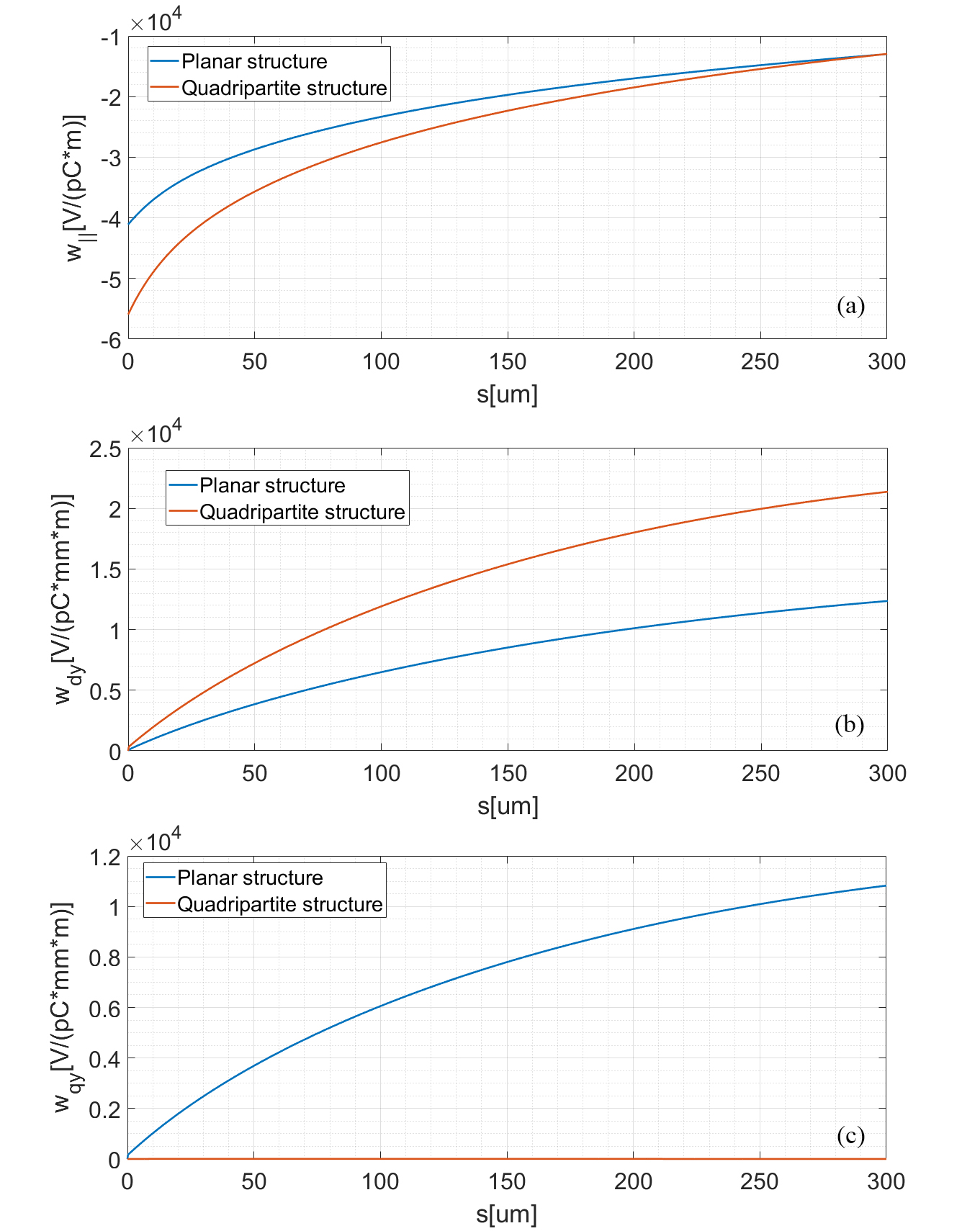}
	\caption{Comparison of one-dimensional wake functions at the nominal effective gap position. (a) The longitudinal wake functions. (b) The dipole wake functions. (c) The quadrupole wake functions.}
	\label{fig_1D_nominal}
\end{figure}

Wakefields of the two structures at the nominal effective gap position are illustrated in Fig.~\ref{fig_1D_nominal}. The longitudinal wakefield and the dipole one are both stronger in the quadripartite structure than the planar one, which can be understood since the shunt impedance of monopole and dipole modes are higher with more confined boundaries. Conversely, the quadrupole in the quadripartite structure is zero, since monopole modes don't contain quadrupole components with the symmetric geometry.

\begin{figure}[!htb]
    \centering
    \includegraphics[width=\hsize]{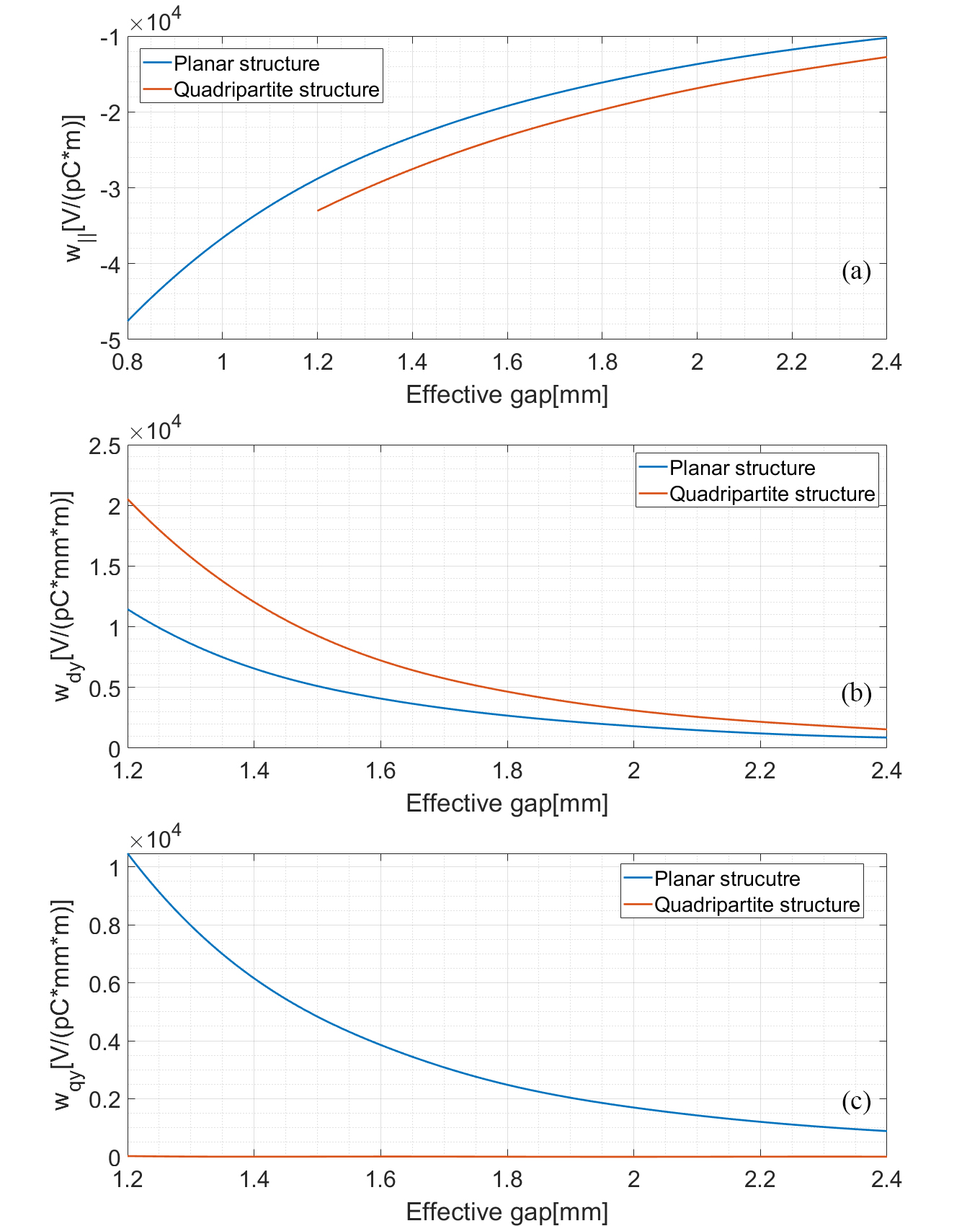}
    \caption{Comparison of one-dimensional wake functions at $s=100~\mu m$ with various effective gaps. (a) The longitudinal wake functions. (b) The dipole wake functions. (c) The quadrupole wake functions.}
    \label{fig_1D_gap}
\end{figure}

The wake functions of the two structures with various effective gaps are illustrated in Fig.~\ref{fig_1D_gap}. The results demonstrate flexibility of the quadripartite structure where the longitudinal wake function varies with the gap while the quadrupole wake remains zero. It should be noted that the maximum available longitudinal wake in the quadripartite structure would be lower than that of the planar structure, which is limited by the minimal effective gap. However, severe projected emittance growth is expected at small effective gaps using the planar structure due to the large quadrupole wakefield.

In the remainder of the manuscript, the nominal effective gap of 1.4~mm is adopted in both structures.

\subsection{Performance for dechirper application}
The use of the two structures as dechirpers is evaluated by \texttt{ELEGANT} with one-dimensional wakefields.

\begin{figure}[!htb]
	\centering
	\includegraphics[width=\hsize]{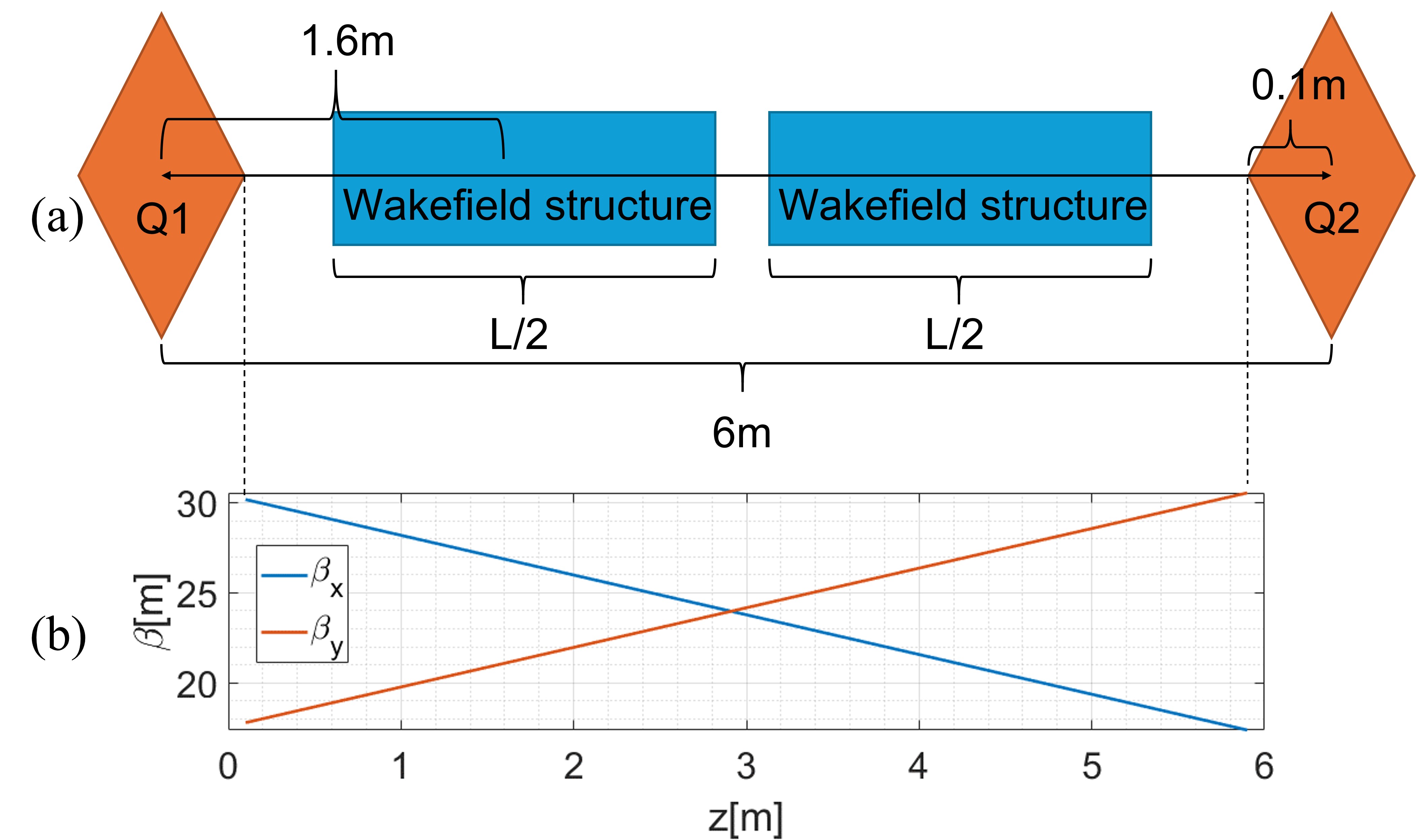}
	\caption{(a) Schematic of the dechirper section in beam dynamics simulation. (b) The $\beta$-functions correspond to the beam parameters in Table~\ref{tab_beam}.}
	\label{fig_dechirpersection}
\end{figure}

The 6~m-long dechirper section consists of two wakefield structures with the same length $L/2$ positioned between two quadrupole magnets (Q1 and Q2), as illustrated in Fig.~\ref{fig_dechirpersection}(a). Each quadrupole magnet has a length of 0.2~m, and the center-to-center distance from each magnet to the nearest wakefield structure is fixed at 1.6~m. For the planar geometry, the two wakefield structures are oriented orthogonally to each other in order to compensate for the quadrupole wakefield~\cite{Zhang_PRAB2015}. In addition, the Twiss parameters are tuned to obtain a nearly identical average $\beta$-function of both transverse direction through the dechirper section, as illustrated in Fig.~\ref{fig_dechirpersection}(b).

Table~\ref{tab_beam} summarizes the beam parameters at the entrance of the dechirper section in this study. The beam is artificially generated from the realistic one at the exit of the main accelerator in S$^{3}$FEL, where the longitudinal phase space, transverse projected emittance, transverse beam size are preserved. However, the generated beam simplifies the transverse phase space by assuming identical Gaussian distribution of particle position, identical emittance, and aligned phase space among slices in order to clearly identify the wakefield effects. Studies with realistic beam distributions will be conducted in future work.

\begin{table}[!htb]
	\centering
	\caption{Beam parameters at the entrance of the dechirper section.}
	\label{tab_beam}
	\begin{ruledtabular}
		\begin{tabular}{l c}
			Parameter & Value \\
			\hline
			Average beam energy & 2544.51~MeV \\
			Pulse charge & 100~pC \\
			Peak current & 880.78~A \\
			Normalized projected emittance $\epsilon_{x}/\epsilon_{y}$ & 0.469/0.414~mm mrad\\
			RMS beam size $\delta_{x}/\delta_{y}$& 53.3/38.5~$\mu$m\\
			RMS bunch length & 52.26~fs (15.67$~\mu$m) \\ 
		\end{tabular}
	\end{ruledtabular}
\end{table}

\begin{figure}[!htb]
    \centering
    \includegraphics[width=\hsize]{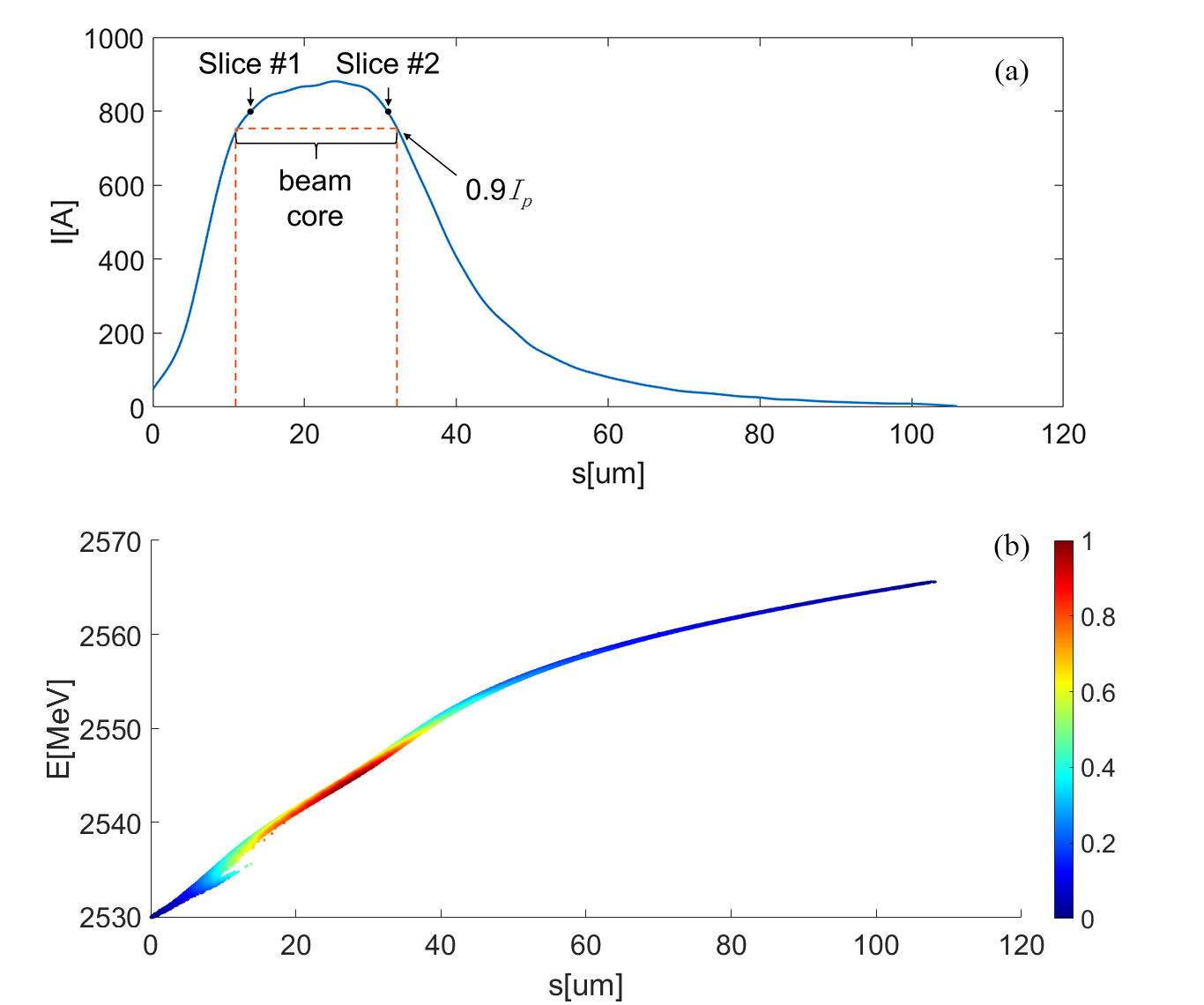}
    \caption{The temporal distribution (a) and the longitudinal phase space (b) of the beam at the entrance of the dechirper section. Two representative slices to be analyzed in Sec.~\ref{beam_size} are marked in (a).}
    \label{fig_beam}
\end{figure}

The temporal distribution is illustrated in Fig.~\ref{fig_beam}(a), where the beam core is defined for the part when the current exceeds 90\% of the peak value. The longitudinal phase space of the initial beam are illustrated in Fig.~\ref{fig_beam}(b), where a large energy chirp can be observed. The chirp is introduced during two-stage magnetic bunch compression and remains uncompensated through the main accelerator to the dechirper section due to the weak wakefield in L-band superconducting cavities with large beam apertures.

\begin{figure}[!htb]
	\centering
	\includegraphics[width=\hsize]{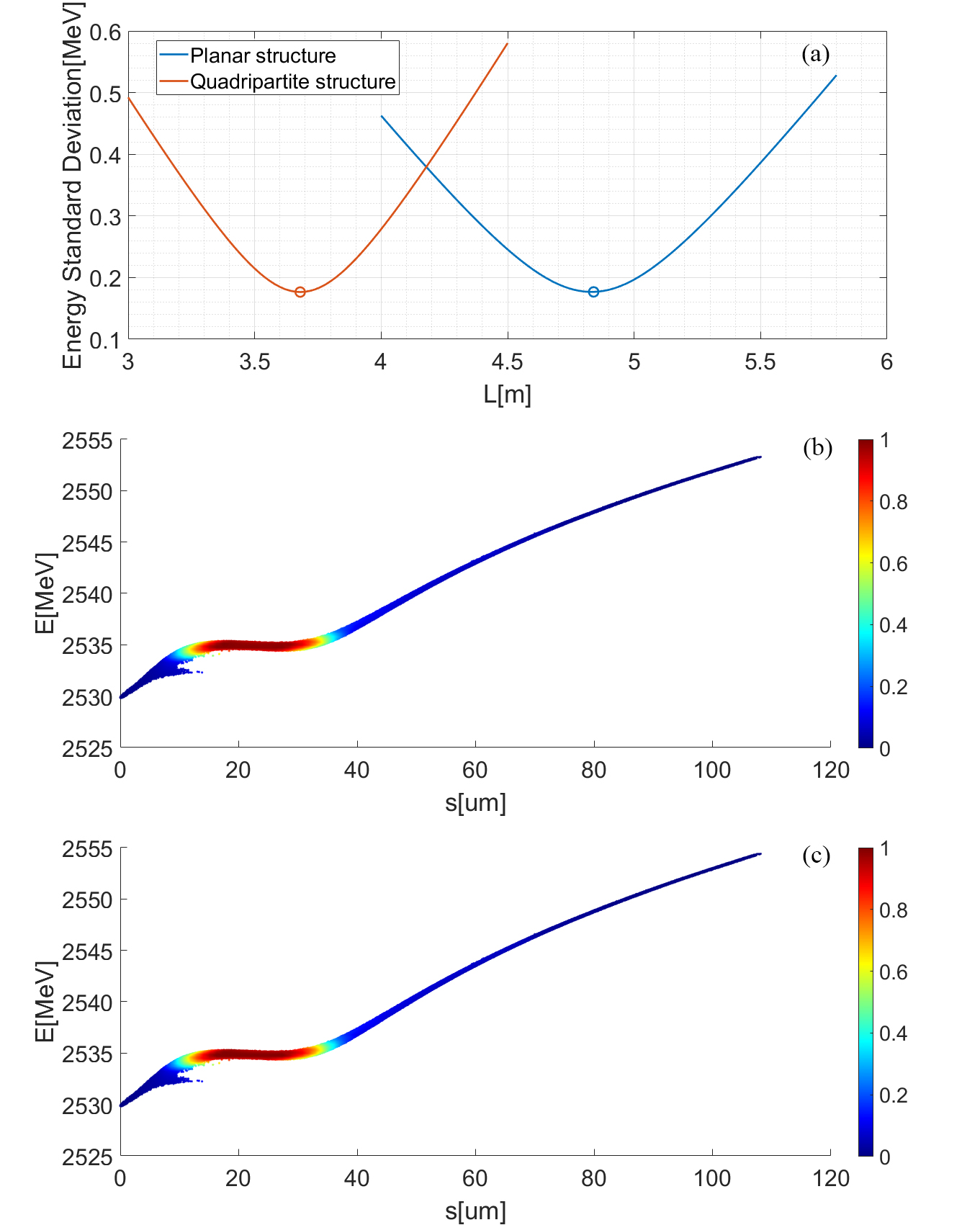}
	\caption{(a) The energy spread of the beam core at the exit of the dechirper section as a function of structure lengths. The minimal values are marked by circles, corresponding to the longitudinal phase spaces of the planar (b) and the quadripartite (c) configurations.}
	\label{fig_dechirper_1D}
\end{figure}

First, the structure lengths are optimized to fully compensate the energy chirp of the beam core. In \texttt{ELEGANT} simulation, the beam traverses on-axis and one-dimensional longitudinal wake functions are applied. As illustrated in Fig.~\ref{fig_dechirper_1D}(a), the energy spread of the beam core at the exit of the dechirper section reduces to a minimum and then grows when the structure length increases, representing scenarios of under-compensation, full-compensation, and over-compensation. The longitudinal phase space with the minimal energy spread is illustrated in Fig.~\ref{fig_dechirper_1D}(b-c). The optimal length of the planar and the quadripartite structures are 4.84~m and 3.66~m, respectively. The ~25\% shorter length of the quadripartite structure results from the stronger longitudinal wake function, as illustrated in Fig.~\ref{fig_1D_nominal}(a). The structure lengths are fixed at the optimal value in following analysis.

Next, the influence on beam projected emittance by the structures is investigated. In \texttt{ELEGANT} simulation, the beam remains on-axis and one-dimensional quadrupole wake functions are added. Two difference configurations are considered for the planar structure: V+H denotes the condition when the vertical and the horizontal structures are respectively placed upstream and downstream, while H+V denotes the reverse one. The vertical orientation is depicted in Fig.~\ref{fig_crosssection}(a) where the corrugation plates move along the vertical direction. 

To evaluate the impact of the time-dependent quadrupole wakefield on slices along the beam, the mismatching factor is defined as follows~\cite{Bettoni_PRAB2015}:
\begin{equation}
	\label{eq_mismatching}
	\Phi (z)=\frac{1}{2}[\beta_{0} \gamma (z)-2 \alpha_{0} \alpha (z)+\gamma_{0} \beta (z)],
\end{equation}
where $\alpha$, $\beta$, and $\gamma$ denote the Twiss parameters along the beam, while $\alpha_{0}$, $\beta_{0}$, and $\gamma_{0}$ denote the Twiss parameters of the projected bunch.

\begin{figure}[!htb]
	\centering
	\includegraphics[width=\hsize]{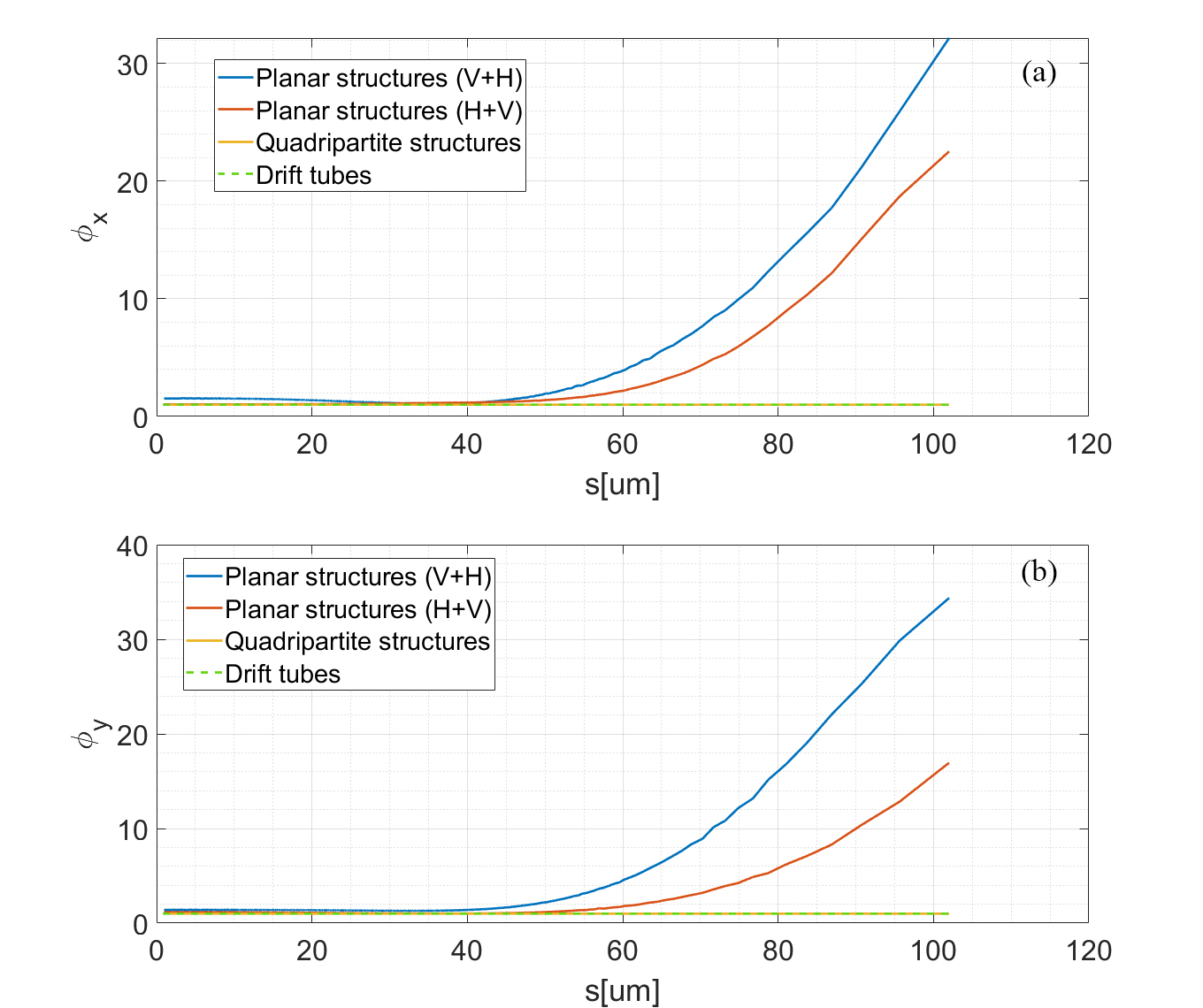}
	\caption{The mismatching factor in the horizontal (a) and the vertical (b) directions for different structure configurations.}
	\label{fig_mismatch}
\end{figure}

Figure~\ref{fig_mismatch} illustrates the mismatching factor along the beam under different structure configurations as well as drifting for reference. It is clear that the mismatching factor grows rapidly along the beam under both planar configurations, indicating the slice transverse phase spaces are rotated by different angles via the time-dependent quadrupole wakefield. Meanwhile, the factor remains close to unity for both drifting and the quadripartite structure, which demonstrates the slices are still well aligned. 

Slice-to-slice mismatch leads to projected emittance growth, which is defined as
\begin{equation}
	\Delta \epsilon(\%)=\frac{\epsilon-\epsilon_{0}}{\epsilon_{0}}
	\label{eq_emittancegrowth}
\end{equation}
where \(\epsilon_0\) and \(\epsilon\) represent the normalized emittance at the entrance and the exit of the dechirper section, respectively. 

The projected emittance growth results are summarized in Table~\ref{tab_emmitance_1D}. Both planar configurations introduces significant emittance growth, while the emittance is preserved using the quadripartite structures.

\begin{table}[!htb]
	\centering
	\caption{Comparison of projected emittance growth under different structure configurations.}
	\label{tab_emmitance_1D}
	\begin{ruledtabular}
	\begin{tabular}{l c c c c}
		\multirow{2}{*}{Structure} & $\epsilon_{x}$ & $\epsilon_{y}$ & $\Delta \epsilon_{x}$ & $\Delta \epsilon_{y}$ \\
		& (mm mrad)& (mm mrad)& (\%) & (\%)\\
		\hline
		Drift & \(0.469\) & \(0.414\) & \(0\) & \(0\) \\
		Planar (V + H) & \(0.841\) & \(0.822\) & \(79.15\) & \(98.43\) \\
		Planar (H + V) & \(0.653\) & \(0.521\) & \(39.17\) & \(25.75\) \\
		Quadripartite & \(0.469\) & \(0.414\) & \(0\) & \(0\) \\
	\end{tabular}
	\end{ruledtabular}
\end{table}

Consequently, it is demonstrated that the quadripartite structure exhibits superior performance for dechirper application, characterized by the short structure length and zero emittance growth which are resulted from the strong longitudinal wakefield and the fully suppressed quadrupole wakefield.

\section{Three-Dimensional Wakefields}\label{sec.V}
In this section, the differences between three-dimensional wakefields and their one-dimensional counterparts are discussed, which necessitates further beam dynamics study using three-dimensional wakefields.

In \texttt{ELEGANT}, the longitudinal wakefield is assumed to be independent of ($x_0, y_0, x, y$), while transverse wakefields are assumed to be linear with the corresponding offsets. Therefore, one-dimensional wakefields applied by \texttt{ELEGANT} can be expressed as follows:
\begin{equation}
	\label{eq_elegant_1D}
	\left\{
	\begin{aligned}	
		W_{\|}(x_{0}, x, y_{0}, y, s)=&W_{\|}( s) \\
		W_{x}(x_{0}, x, y_{0}, y, s)=&W_{dx}(s)x_{0}+W_{qx}(s)x \\
		W_{y}(x_{0}, x, y_{0}, y, s)=&W_{dy}(s)y_{0}+W_{qy}(s)y \\
	\end{aligned}
	\right.,
\end{equation}

\begin{figure*}[!htb]
	\centering
	\includegraphics[width=\hsize]{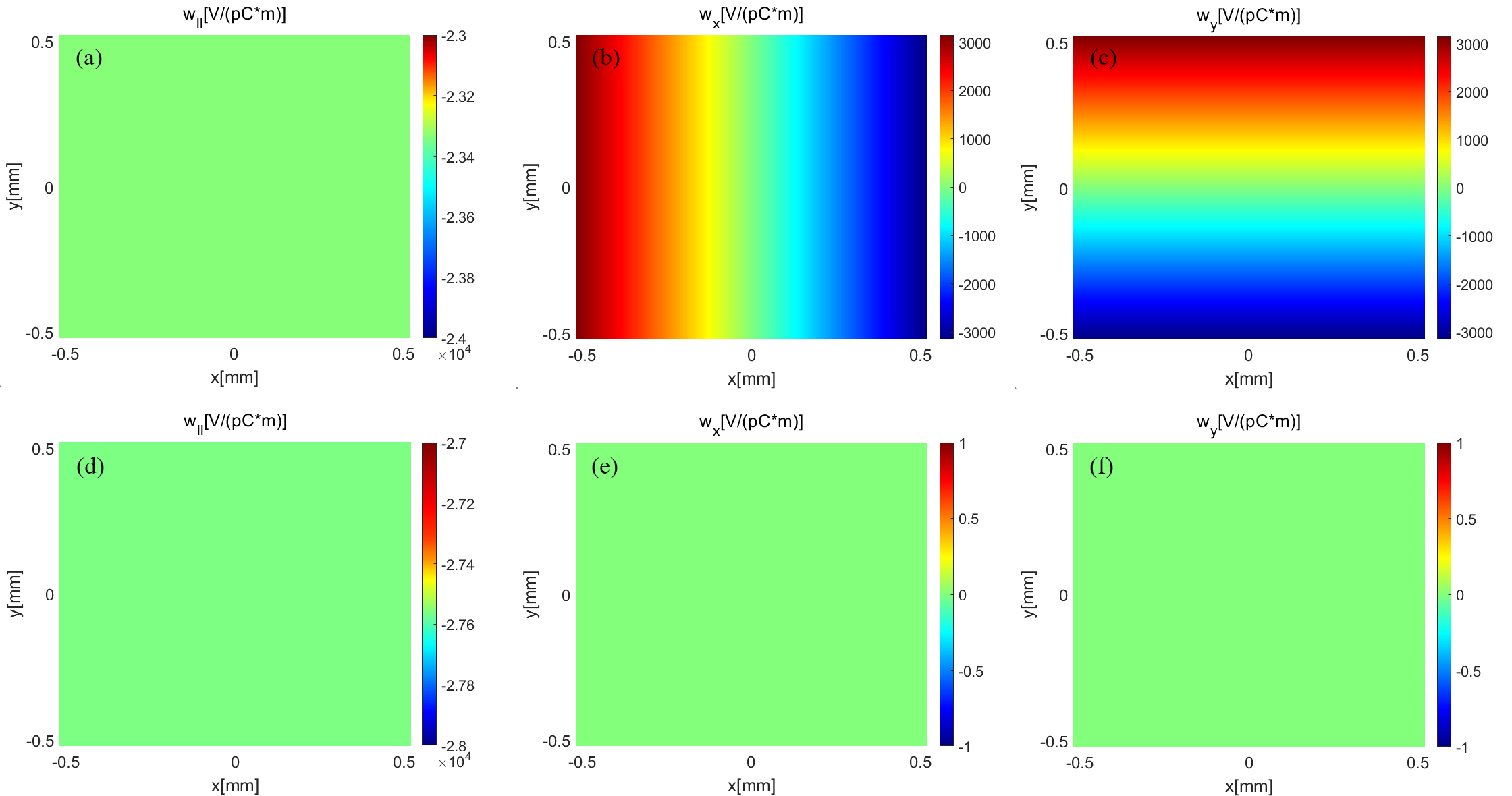}
	\caption{The distributions of ideal three-dimensional wake functions of the planar (top) and the quadripartite (bottom) structures. Left to right: longitudinal, horizontal, and vertical wake functions.}
	\label{fig_3D_ideal}
\end{figure*}

Accordingly, the ideal three-dimensional wakefields can be reconstructed. An example of wakefield distributions with $x_0=y_0=0$ and $s$=100~$\mu$m for the two structures is illustrated in Fig.~\ref{fig_3D_ideal}.

\begin{figure*}[!htb]
	\centering
	\includegraphics[width=\hsize]{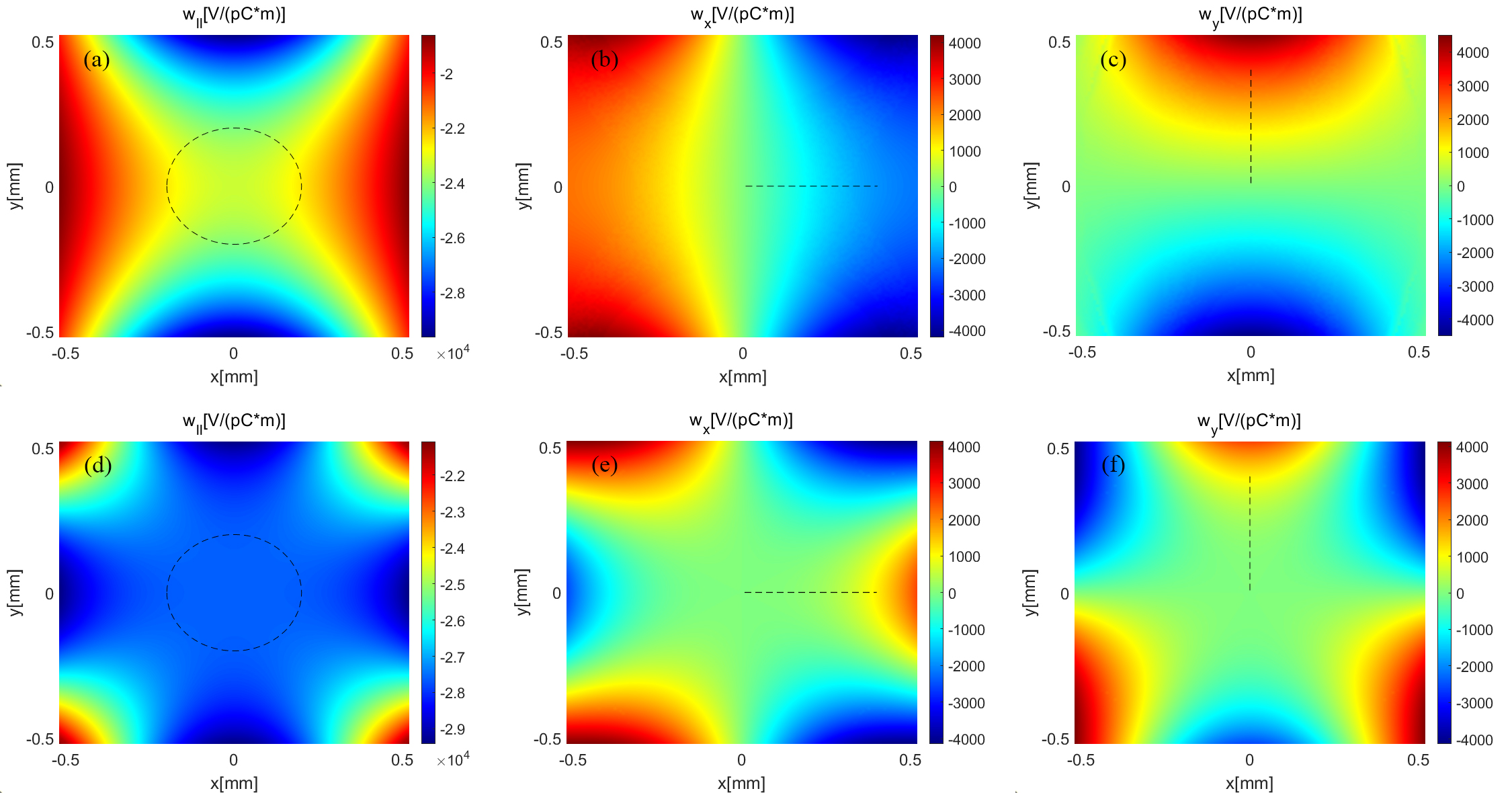}
	\caption{The distributions of realistic three-dimensional wake functions of the planar (top) and the quadripartite (bottom) structures. Left to right: longitudinal, horizontal, and vertical wake functions.}
	\label{fig_3D_realistic}
\end{figure*}

Conversely, the realistic three-dimensional wakefields can be calculated from Eq.~\eqref{eq_PW}. The wakefield distribution for the two structures, with the same coordinates as in Fig.~\ref{fig_3D_ideal}, are illustrated in Fig.~\ref{fig_3D_realistic}.

Compared with the ideal distribution, two remarkable features can be observed in the realistic wakefield distribution. First, the longitudinal wakefield function is not constant for arbitrary $(x, y)$ position. The azimuthal variation can be expressed by the sum of a series of higher-order components as follows~\cite{Chae_PRAB2011}:
\begin{equation}
	\begin{aligned}
		\label{eq_azimuthal}
		W_{||}(x_{0}, x, y_{0}, y, s)=\sum_{n=0}^{\infty} W_{n}(r_0, r, s)cos(n\theta),
	\end{aligned}
\end{equation}
where $W_{n}$ denotes the amplitude of $n$th-order component, $r_0=\sqrt{x_0^2+y_0^2}$, $r=\sqrt{x^2+y^2}$, and $\theta$ denotes the angle between $(x_0, y_0)$ and $(x, y)$ in the polar coordinate. 

\begin{figure}[!htb]
	\centering
	\includegraphics[width=\hsize]{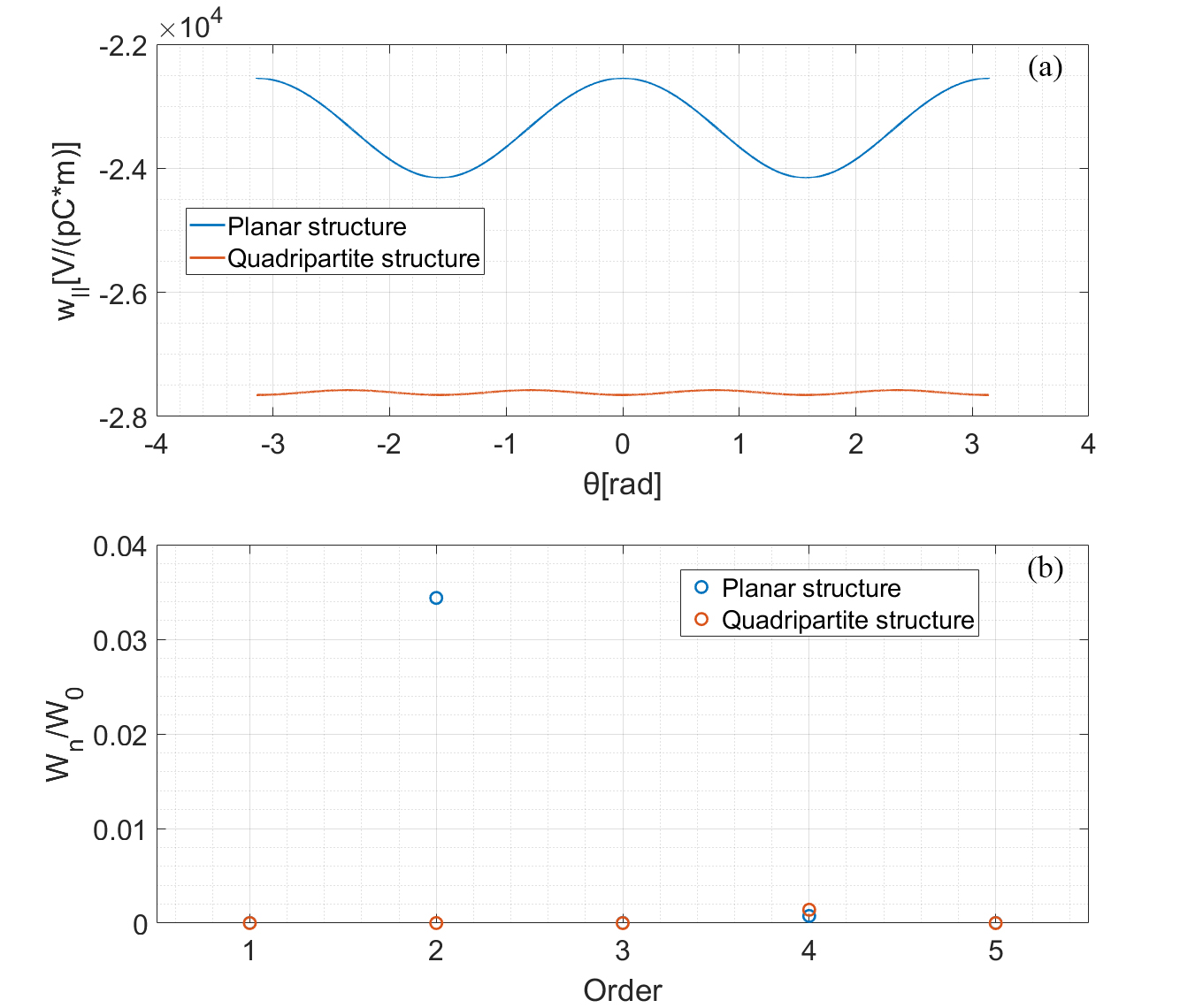}
	\caption{(a) The longitudinal wake function along the azimuthal direction. (b) The amplitude of the higher-order components normalized by the $0$th-order one.}
	\label{fig_azimuthal}
\end{figure}

The longitudinal wakefields along the dashed circles ($r_0$=0, $r$=0.2~mm) in Fig.~\ref{fig_3D_realistic}(a) and (d) are illustrated in Fig.~\ref{fig_azimuthal}(a), while the resultant amplitudes of higher-order components normalized by the 0th-order strength are illustrated in Fig.~\ref{fig_azimuthal}(b). As expected, the quadrupole component ($n$=2) is dominating in the planar structure, but fully suppressed in the quadripartite counterpart. Additionally, octupole components ($n$=4) can be found in both structures, since neither of them is octupole symmetric. The octupole components of the planar structure is slightly lower than the quadripartite structure owing to the larger corrugation width $w$. Further quantitative analysis of the dependence of higher-order components on $w$ is provided in Appendix~\ref{structruewidth}.

\begin{figure}[!htb]
	\centering
	\includegraphics[width=\hsize]{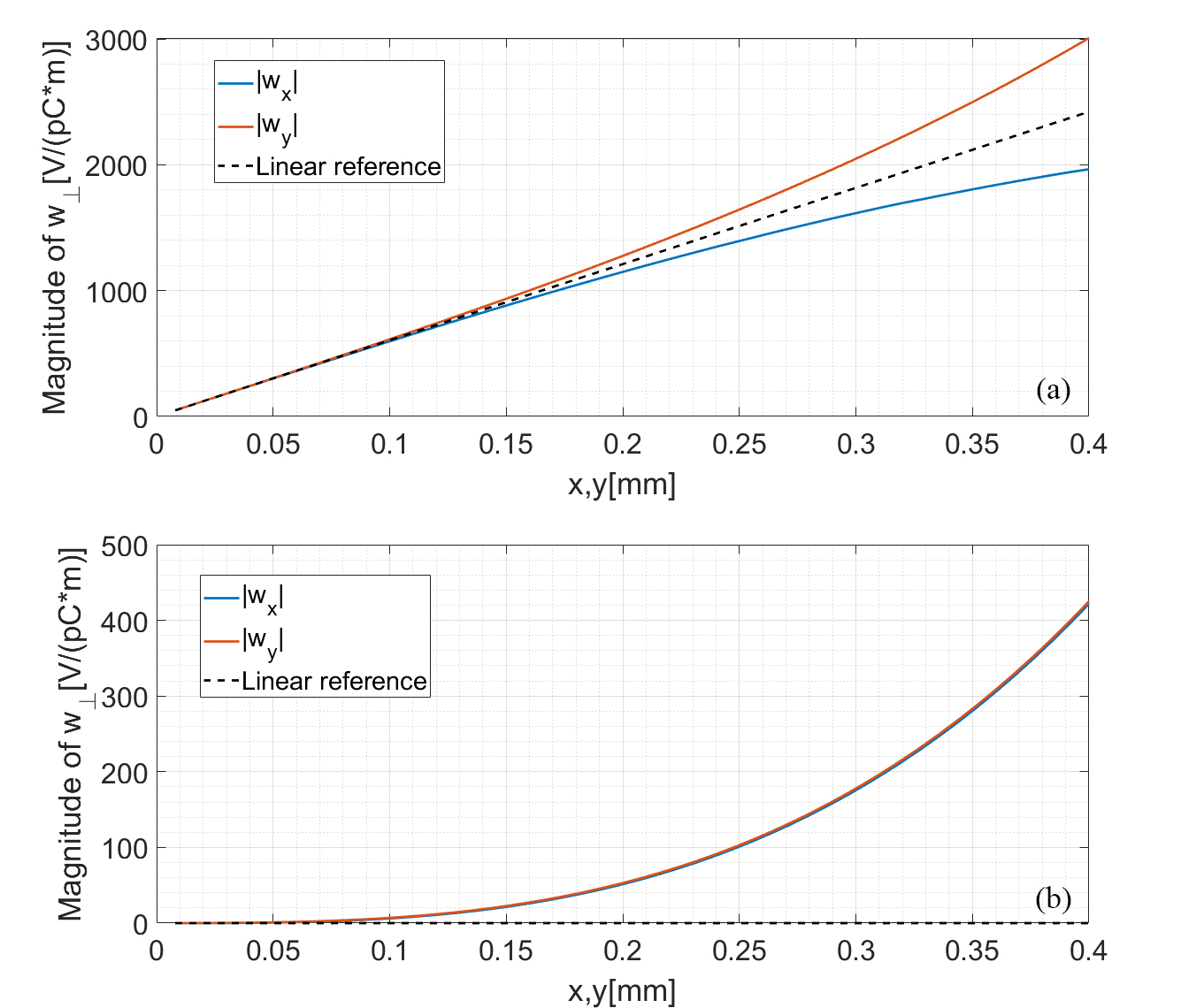}
	\caption{The transverse wake function along the radial directions for the planar (a) structure and the quadripartite (b) structure. The linear dependence in one-dimensional wakefield is provided for reference. }
	\label{fig_radial}
\end{figure}

Second, the transverse wakefield is not linear with the corresponding coordinate. For example,  the transverse wakefields along the dashed lines in radial directions in Fig.~\ref{fig_3D_realistic}(b-c) and (e-f) are illustrated in Fig.~\ref{fig_radial}. For the vertical planar structure, $w_y$ is stronger than the linear approximation with large off-axis while $w_x$ behaves oppositely. For the quadripartite structure, both $w_x$ and $w_y$ show strong non-linearity. The nonlinear effects indicate that higher-order terms in the Taylor expansion in Eq.~\eqref{eq_Taylor} are not negligible for large $(x, y)$. 

Therefore, it is necessary to conduct beam dynamics simulation with three-dimensional wakefields to investigate the impact of higher-order components and nonlinear effects.

\section{Structure comparison with three-dimensional wakefields}\label{sec.VI}
\subsection{Particle-to-particle method}
A framework based on the particle-to-particle method~\cite{Wang_Thesis} is developed in order to apply three-dimensional wakefields. The force experienced by any particle (denoted as the witness particle) within the bunch is determined by the cumulative contributions from all preceding particles (denoted as driving particles) as follows
\begin{equation}
	\label{eq_p2p}
	\left\{
	\begin{aligned}
		F_{||}(x_{m}, y_{m}, s_m)=&q^2 \sum_{i} w_{||}(x_{i}, x_{m}, y_{i}, y_{m}, s_{i}-s_{m})\\
		F_{x}(x_{m}, y_{m}, s_m)=&q^2 \sum_{i} w_{x}(x_{i}, x_{m}, y_{i}, y_{m}, s_{i}-s_{m})\\
		F_{y}(x_{m}, y_{m}, s_m)=&q^2 \sum_{i} w_{y}(x_{i}, x_{m}, y_{i}, y_{m}, s_{i}-s_{m})\\
	\end{aligned}
	\right.,
\end{equation}
where $q$ is the particle charge, and ($x_m$, $y_m$, $s_m$) and ($x_i$, $y_i$, $s_i$) denote the positions of the $m$th witness particle and the $i$th driving particle, respectively.

Although \texttt{ECHO3D} is capable to provide wakefields at arbitrary positions for a given beam offset in a single simulation, performing such calculations for every driving particle at each time step along the dechirper section would require an excessive number of simulations. Considering that the numbers of driving particles and time steps are on the order of $1\times10^4$ and $1\times10^2$ respectively, approximately $1\times10^6$ wakefield simulations would be needed, which is computationally impractical. Therefore, a simplified approach is used where the drive particle offsets are averaged and the resultant three-dimensional wakefields of the single offset is then applied to the witness particle as
\begin{equation}
	\label{eq_p2p_simplified}
	\left\{
	\begin{aligned}		
		F_{||}(x_{m}, y_{m}, s_w)=&q^2 \sum_{i} w_{||}(\overline{x}, x_{m}, \overline{y}, y_{m}, s_{i}-s_{m})\\
		F_{x}(x_{m}, y_{m}, s_w)=&q^2 \sum_{i} w_{x}(\overline{x}, x_{m}, \overline{y}, y_{m}, s_{i}-s_{m})\\
		F_{y}(x_{m}, y_{m}, s_w)=&q^2 \sum_{i} w_{y}(\overline{x}, x_{m}, \overline{y}, y_{m}, s_{i}-s_{m})\\
	\end{aligned}
	\right.,
\end{equation}
where ($\overline{x}$, $\overline{y}$) are the average offset of the entire bunch.

The simplified approach relies on several assumptions as follows. 

First, the average wakefields of each particle can be approximated by the wakefields of the average offset:
\begin{equation}
	\label{eq_p2p_approximation}
	\left\{
	\begin{aligned}
		\frac{1}{N}\sum_{i} w_{||}\left(x_{i}, x, y_{i}, y, s\right) \approx w_{||}\left(\overline{x}, x, \overline{y}, y, s\right)\\
		\frac{1}{N}\sum_{i} w_{x}\left(x_{i}, x, y_{i}, y, s\right) \approx w_{x}\left(\overline{x}, x, \overline{y}, y, s\right)\\
		\frac{1}{N}\sum_{i} w_{y}\left(x_{i}, x, y_{i}, y, s\right) \approx w_{y}\left(\overline{x}, x, \overline{y}, y, s\right)
	\end{aligned}
	\right.,
\end{equation}
where $N$ is the particle number in the bunch. This approximation depends on the ratio between the beam size and the structure aperture and may become inaccurate when this ratio is large. For the structures considered in this study, and for a beam with dimensions twice those listed in Table~\ref{tab_beam}, the approximation remains valid, as demonstrated by the detailed analysis in Appendix~\ref{app}. 

Second, the average offset is assumed to remain constant along the structure. This assumption holds for the current study, where the beam propagates on-axis with a symmetric transverse distribution in both horizontal and vertical planes. For beams exhibiting significant transverse asymmetry or large off-axis trajectories, focusing and defocusing effects would cause the average offset to vary along the structure. In such cases, the structure should be divided into multiple segments (on the order of $1\times10^1$), within which the average offset can be treated as constant, thereby requiring additional \texttt{ECHO3D} simulations for each segment.

Third, the average offset is constant among slices along the bunch, which is valid for a well-aligned beam. For a largely tilted beam, the method could be extended to account for the wakefields generated by the averaged position of each slice. When both the beam is divided into slices (on the order of $1\times10^1$) and the structure into segments, approximately $1\times10^2$ \texttt{ECHO3D} simulations would be needed, which is still computationally practical. The modification related to these beam conditions will be addressed in future studies.

Fourth, beam loss does not change the average offset of the beam.

Based on this simplified approach, the particle-to-particle method is implemented by \texttt{MATLAB} codes and the algorithm workflow can be found in Ref.~\cite{Wang_Thesis}. Different from the approach in Ref.~\cite{Wang_Thesis} which uses only a few wakefields modes calculated by RF properties of resonant cavities, the method in this study imports the realistic three-dimensional wake functions from \texttt{ECHO3D} simulation and deconvolution that include the aforementioned higher-order components and nonlinear effects. 

To validate the particle-to-particle method, a comparison against \texttt{ELEGANT} simulations is conducted using the ideal wakefield distribution introduced in Eq.~\eqref{eq_elegant_1D} and illustrated in Fig.~\ref{fig_3D_ideal}. The normalized projected emittance at the end of the dechirper section with planar structure under H+V configuration is summarized in  Table~\ref{tab_emittance verification}, where the negligible differences confirms the reliability of the particle-to-particle method.

\begin{table}[!htb]
	\centering
	\caption{Comparison of projected emittance between different methods.}
	\label{tab_emittance verification}
	\begin{ruledtabular}
		\begin{tabular}{l c c}
			\multirow{2}{*}{Method} & $\epsilon_{x}$  & $\epsilon_{y}$  \\
			& (mm mrad) & (mm mrad) \\
			\hline
			\texttt{ELEGANT} & 0.653 & 0.521 \\
			Particle-to-particle & 0.654 & 0.519\\
			Difference & 0.21\% & -0.48\%\\
		\end{tabular}
	\end{ruledtabular}
\end{table}

\subsection{Simulation results with the nominal beam size}
The performance of the structures for dechirper applications is re-evaluated with the particle-to-particle approach. The same structure parameters of \texttt{ELEGANT} simulations in Sec.~\ref{sec.IV} are applied, where the effective gap is 1.4~mm and the total lengths are 4.84~m and 3.66~m for the planar and quadripartite structure, respectively. The nominal beam parameters as listed in Table~\ref{tab_beam} is used and the lattice illustrated in Fig.~\ref{fig_dechirpersection} is adopted.

\begin{figure*}[!htb]
	\includegraphics[width=\hsize]{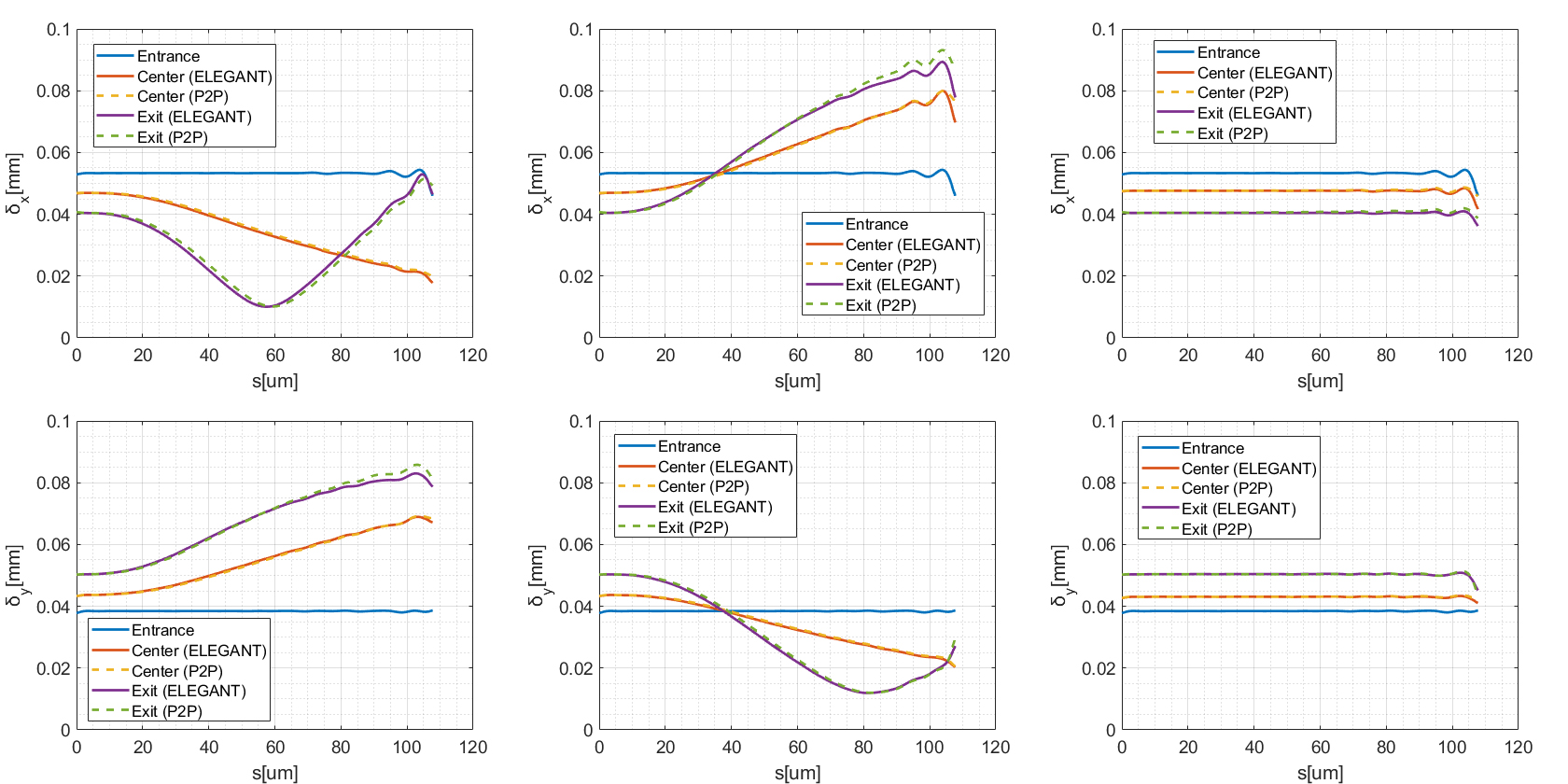}
	\caption{The RMS slice beam size at the entrance, center, and exit of the dechirper section. Top: horizontal plane; bottom: vertical plane. Left to right: planar structures (V+H), planar structures (H+V), quadripartite structures.}
	\label{fig_tracking_beamsize}
\end{figure*}

The RMS slice beam size along the dechirper section is illustrated in Fig.~\ref{fig_tracking_beamsize}. For the quadripartite structures, the slice beam size distributions remain nearly constant during beam propagation, and their evolution follows the beta function in Fig.~\ref{fig_dechirpersection}(b), confirming the effective suppression of quadrupole wakefields. In contrast, pronounced variations in slice beam size along the bunch are observed for the planar structures in both configurations, indicating strong time-dependent quadrupole wakefields excited within the structures. Such large slice-to-slice differences in beam size could pose challenges for stable beam transport under realistic operating conditions.

\begin{figure*}[!htb]
	\includegraphics[width=\hsize]{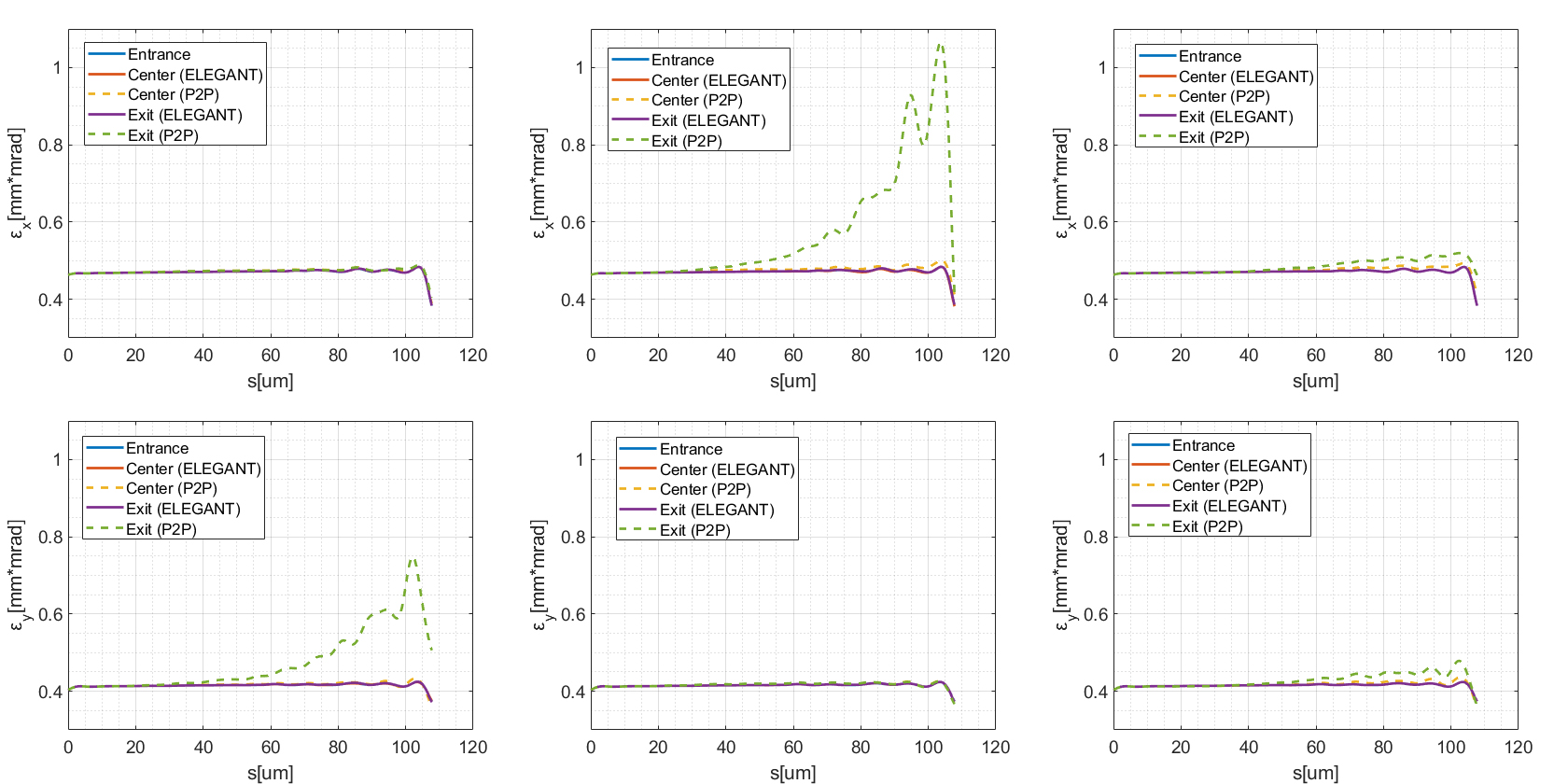}
	\caption{The normalized slice emittance at the entrance, center, and exit of the dechirper section. Top: horizontal plane; bottom: vertical plane. Left to right: planar structures (V+H), planar structures (H+V), quadripartite structures.}
	\label{fig_tracking_emittance}
\end{figure*}

The evolution of the normalized slice emittance along the dechirper section is illustrated in Fig.~\ref{fig_tracking_emittance}. In contrast to the close agreement observed for the slice beam size between the \texttt{ELEGANT} tracking and the particle-to-particle approach, noticeable discrepancies are observed in the slice emittance. In \texttt{ELEGANT} simulations, the slice emittance is well preserved during propagation because the applied one-dimensional transverse wakefields include only linear terms. Conversely, the three-dimensional transverse wakefields used in the particle-to-particle approach contain nonlinear components that induce emittance growth within individual slices~\cite{Chae_PRAB2011}. The effect is most prominent in the tail slices when using the planar structures, where the wakefield nonlinearities become more significant for larger slice beam sizes (Fig.~\ref{fig_radial}.)

\begin{figure}[!htb]
	\centering
	\includegraphics[width=\hsize]{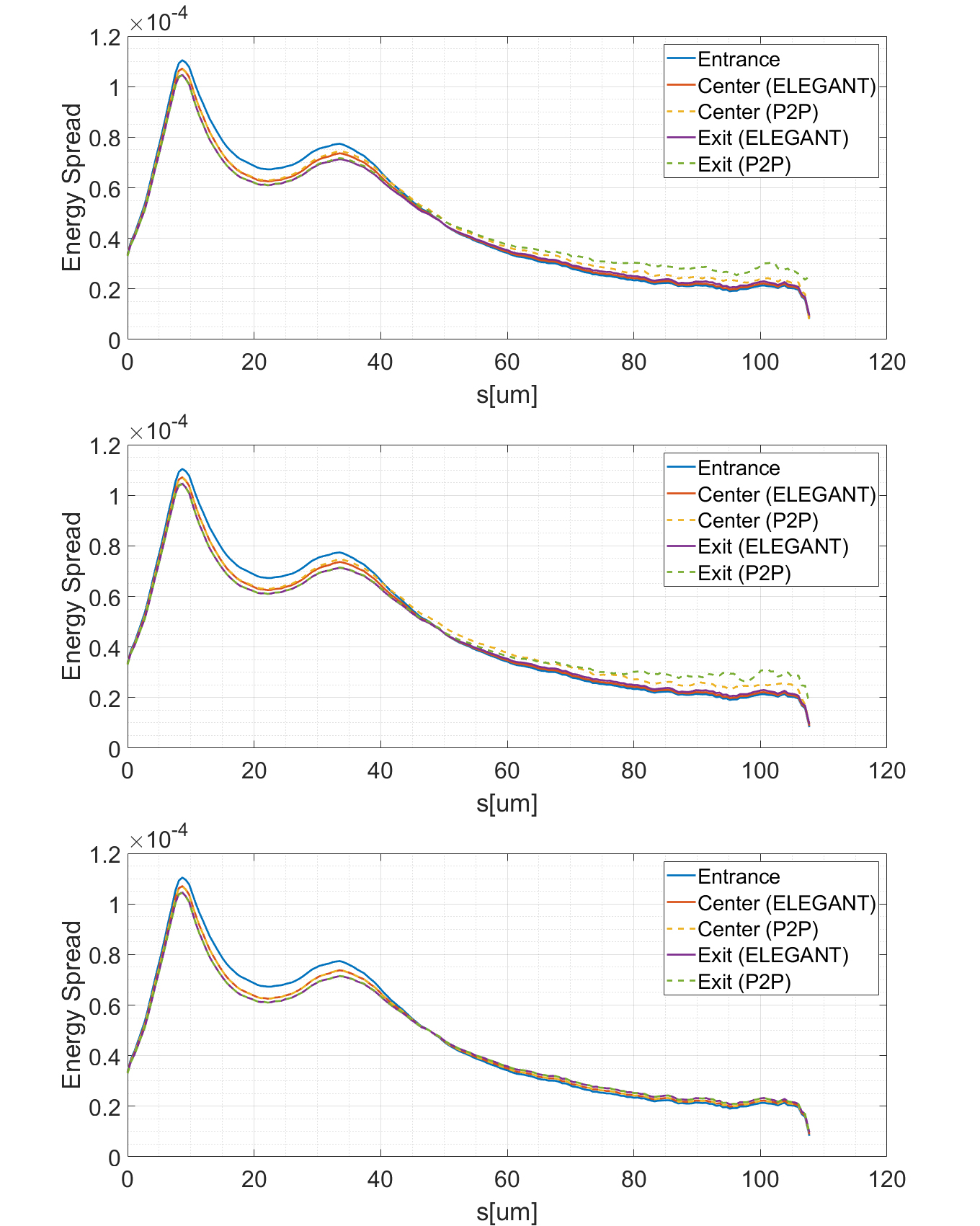}
	\caption{The slice energy spread at the entrance, center, and exit of the dechirper section. Top to bottom: planar structures (V+H), planar structures (H+V), quadripartite structures.}
	\label{fig_tracking_energyspread}
\end{figure}

Similarly, the influence of three-dimensional wakefields is evident in the evolution of the slice energy spread, as illustrated in Fig.~\ref{fig_tracking_energyspread}. Because the longitudinal wakefield varies across the transverse plane, particularly in the planar structures (Fig.~\ref{fig_3D_realistic}), a finite beam size introduces additional slice energy spread during propagation.

The normalized projected emittance at the exit of the dechirper section is summarized in Table~\ref{tab_emittance_1D_3D}. The particle-to-particle method confirms significant emittance growth with the planar structure under both configurations, while shows negligible emittance growth for the quadripartite structure even when higher-order components and nonlinear effects from three-dimensional wakefields are taken into account.

\begin{table*}[!htb]
	\centering
	\caption{Comparison of projected emittance growth under different structure configurations and simulation methods.}
	\label{tab_emittance_1D_3D}
	\begin{ruledtabular}
		\begin{tabular*}{16cm} {@{\extracolsep{\fill} } llcccc}
			Structure & Simulation method & $\epsilon_{x}$ (mm mrad) & $\epsilon_{y}$ (mm mrad) & $\Delta \epsilon_{x} (\%)$ & $\Delta \epsilon_{y} (\%)$ \\
			\hline
			\multirow{2}{*}{Planar structures (V+H)}& \texttt{ELEGANT} & 0.841 & 0.822 & 79.15 & 98.43 \\ 
			& Particle-to-particle& 0.815 & 0.770 & 73.59 & 85.87 \\
			\multirow{2}{*}{Planar structures (H+V)}& \texttt{ELEGANT} & 0.653 & 0.521 & 39.17 & 25.75 \\
			& Particle-to-particle& 0.618 & 0.519 & 31.54 & 25.41 \\
			\multirow{2}{*}{Quadripartite structures} & \texttt{ELEGANT} & 0.469 & 0.414 & 0 & 0 \\
			& Particle-to-particle& 0.471 & 0.415 & 0.23 & 0.31 \\  
		\end{tabular*}
	\end{ruledtabular}
\end{table*}

Self-amplified spontaneous emission (SASE)~\cite{Kondratenko_PA1980,Bonifacio_OC1984} at 1~nm under these beam conditions is simulated with the \texttt{GENESIS} 1.3 code~\cite{Reiche_NIMA1999}. A total of 27 undulator segments are included to ensure full saturation and to capture the post-saturation behavior~\cite{Yi_PRAB2025}. Each undulator segment is 4~m long and contains 133 undulator periods with a 30~mm period length. The undulator segments are separated by 1-meter-long break sections equipped with quadrupoles to maintain the FODO structure. Since the mismatching factor, slice emittance, and energy spread of the beam core remain nearly unchanged for both the planar and quadripartite structures when employing either one-dimensional or three-dimensional wakefields, no statistically significant differences are observed in the resulting lasing performance.

Because the V+H planar configuration leads to significantly larger projected emittance growth than the H+V case, only the latter setup is retained for comparison with the quadripartite structure in the following analysis.

\subsection{Simulation results with various beam sizes}\label{beam_size}
To further evaluate the influence of higher-order components and nonlinear effects in three-dimensional wakefields, the nominal beam sizes in Table~\ref{tab_beam} are scaled by a factor of 0.7-2.0 (denoted as the beam size ratio) for additional tracking and lasing simulations. The minimal size is determined by the beam emittance and the maximum quadrupole strength, while the maximum size is limited by noticeable beam loss inside the structures. For each beam size, the lattice is re-optimized to ensure a nearly identical average $\beta$-function of both transverse planes along the dechirper section.

\begin{figure}[!htb]
	\centering
	\includegraphics[width=\hsize]{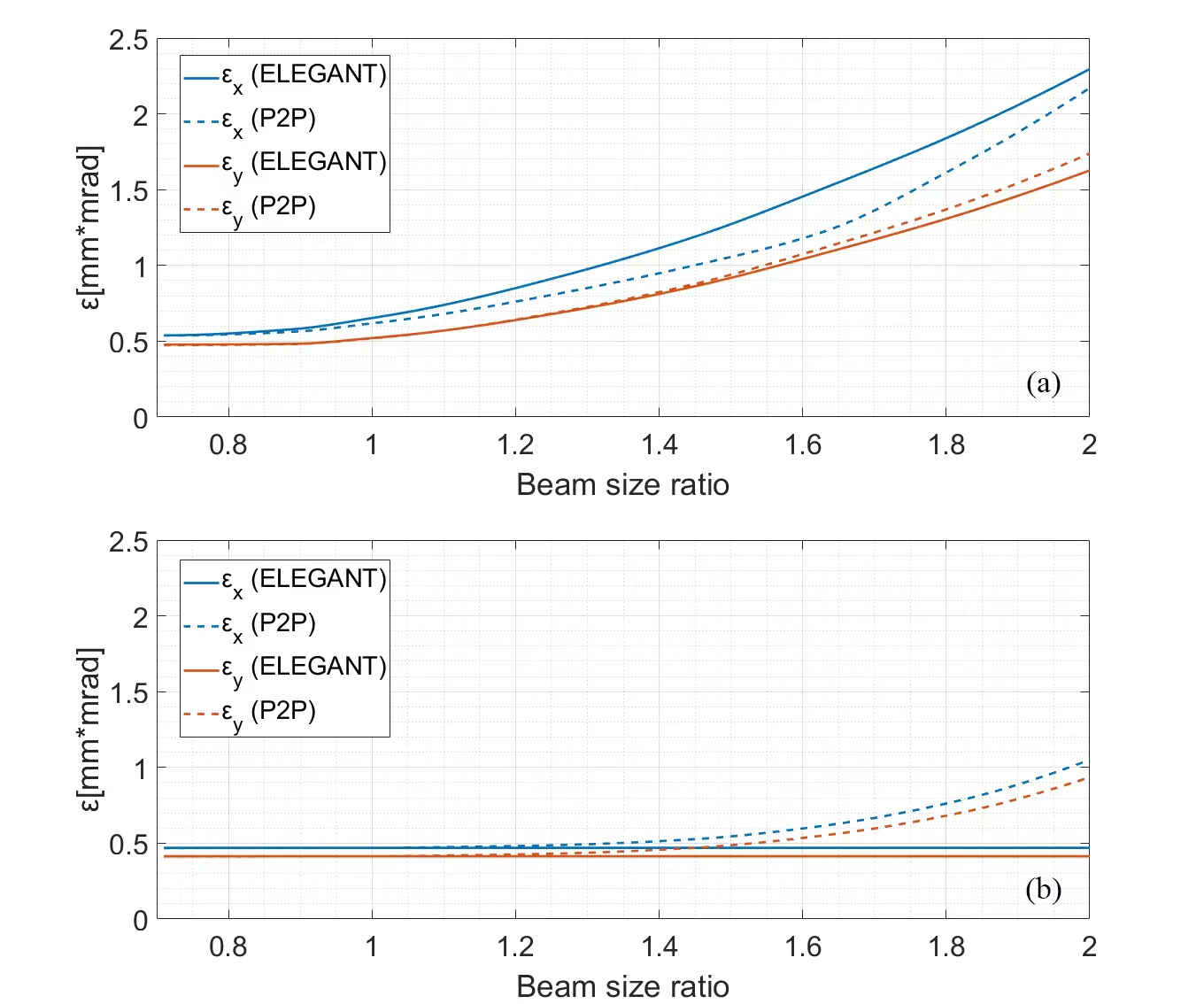}
	\caption{The normalized projected emittance at the exit of the dechirper section as a function of the beam size ratio. (a) Planar structures with H+V configuration. (b) Quadripartite structures.}
	\label{fig_projectedemittance_beamsize}
\end{figure}

\begin{figure}[!htb]
	\centering
	\includegraphics[width=\hsize]{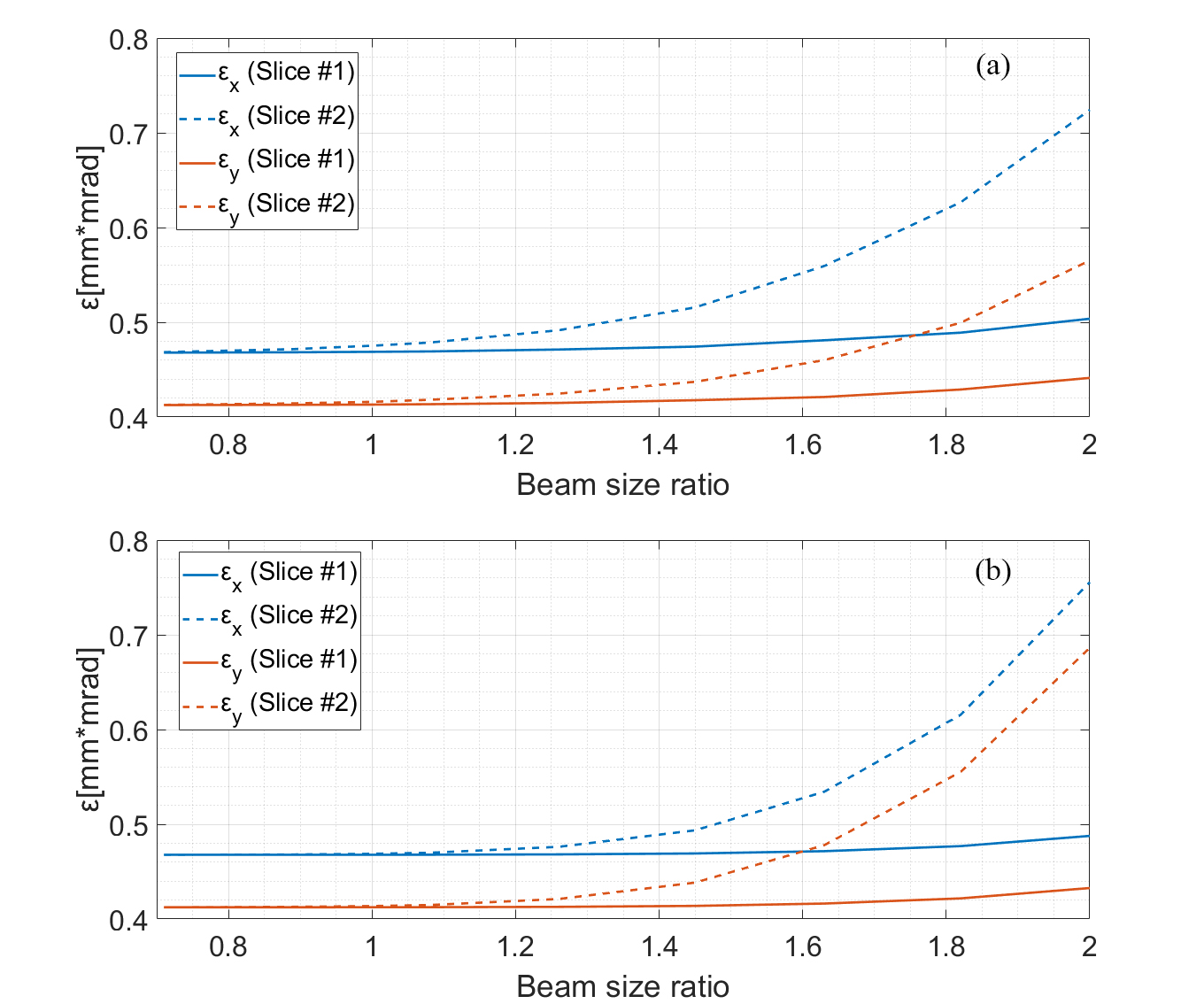}
	\caption{The normalized slice emittance of the two slices indicated in Fig.~\ref{fig_beam} at the exit of the dechirper section as a function of the beam size ratio (particle-to-particle results only). (a) Planar structures with H+V configuration. (b) Quadripartite structures.}
	\label{fig_sliceemittance_beamsize}
\end{figure}

With various beam sizes, the normalized projected emittance of the entire beam and the normalized slice emittance of two slices with beam current of 800~A in the beam core (Fig.~\ref{fig_beam}) at the exit of the dechirper section are illustrated in Fig.~\ref{fig_projectedemittance_beamsize} and Fig.~\ref{fig_sliceemittance_beamsize}, respectively.

Slice emittance growth can be observed for both structures in the particle-to-particle tracking, particularly for larger transverse beam sizes, due to the nonlinear components of the three-dimensional wakefields. In the quadripartite structure, the accumulated slice emittance growth ultimately leads to a noticeable increase in the projected emittance. In contrast, for the planar structure, the influence of slice emittance growth on the projected emittance is limited, indicating that the latter is dominated by slice-to-slice mismatch driven by the quadrupole-wakefield component.

\begin{figure}[!htb]
	\centering
	\includegraphics[width=\hsize]{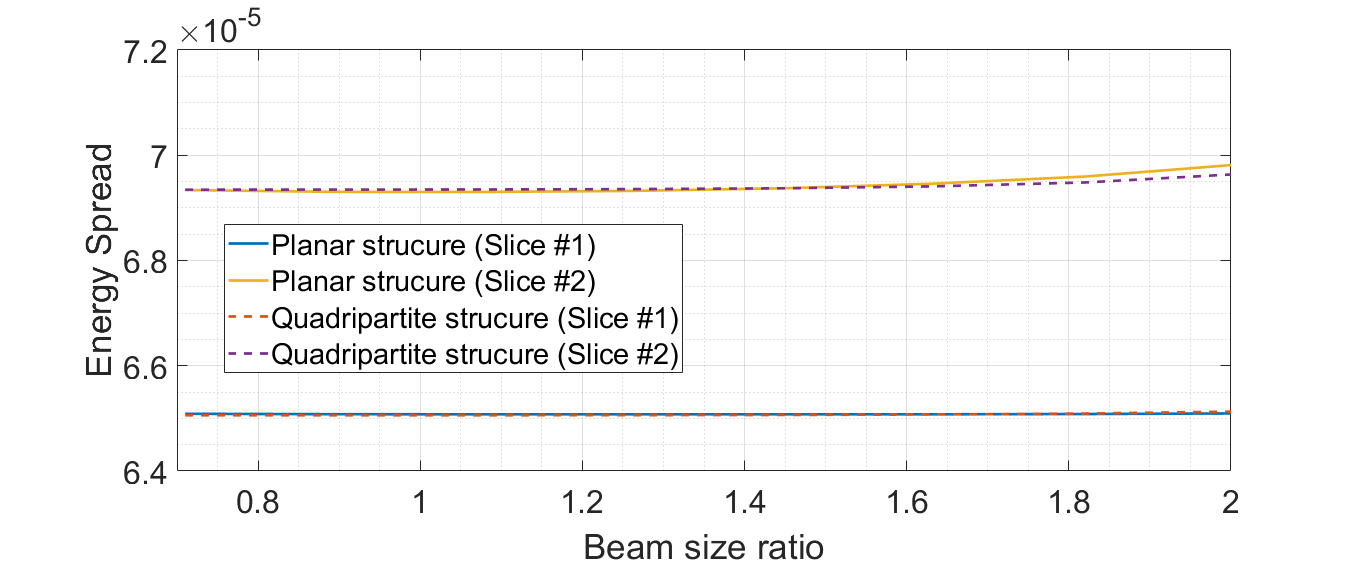}
	\caption{The slice energy spread of the two slices indicated in Fig.~\ref{fig_beam} at the exit of the dechirper section as a function of the beam size ratio (particle-to-particle results only).}
	\label{fig_sliceenergyspread_beamsize}
\end{figure}

Meanwhile, the impact of the beam size on the slice energy spread is relatively modest, as illustrated in Fig.~\ref{fig_sliceenergyspread_beamsize}.

\begin{figure}[!htb]
	\centering
	\includegraphics[width=\hsize]{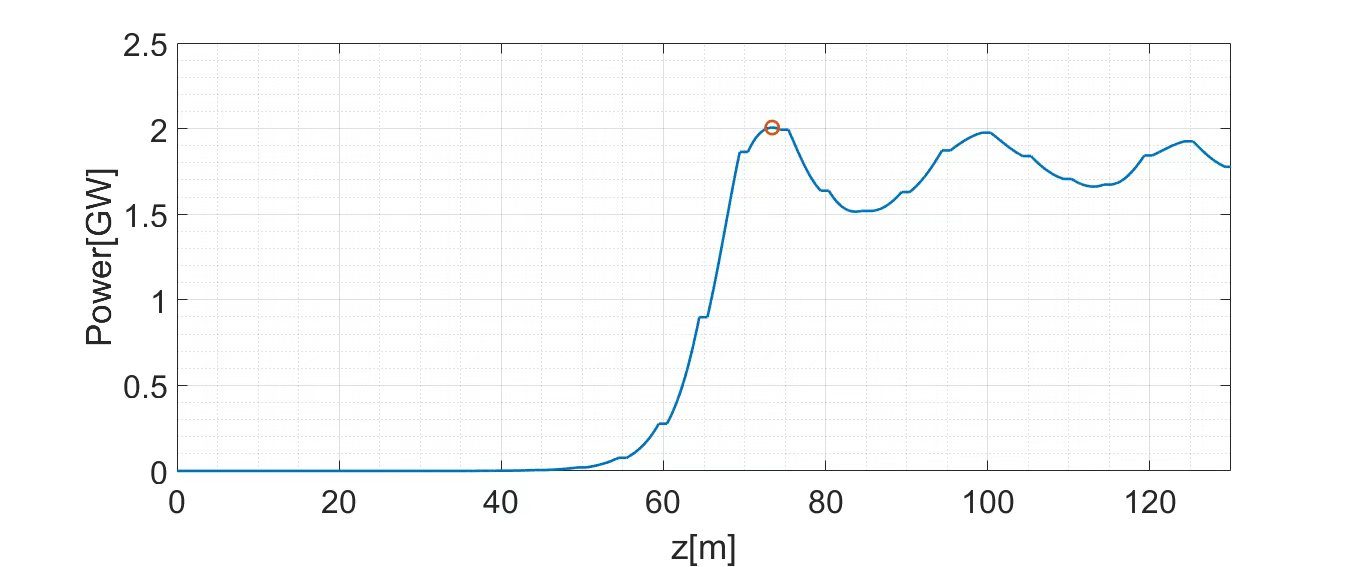}
	\caption{Evolution of the peak SASE radiation power at 1~nm along the undulators for quadripartite structures using the smallest beam size. The saturation point is marked by the red circle.}
	\label{fig_lasingcurve}
\end{figure}

To assess the impact of slice emittance degradation, lasing at 1~nm is simulated in the \texttt{GENESIS} 1.3 code using Slice \#2 instead of the entire bunch. A representative evolution of radiation power along the undulators is illustrated in Fig.~\ref{fig_lasingcurve}. The saturation point, defined as the transition from exponential growth to the maximum output power, is clearly observed. Beyond this point, the power exhibits the characteristic decline and oscillatory behavior of the post-saturation regime.

\begin{figure}[!htb]
	\centering
	\includegraphics[width=\hsize]{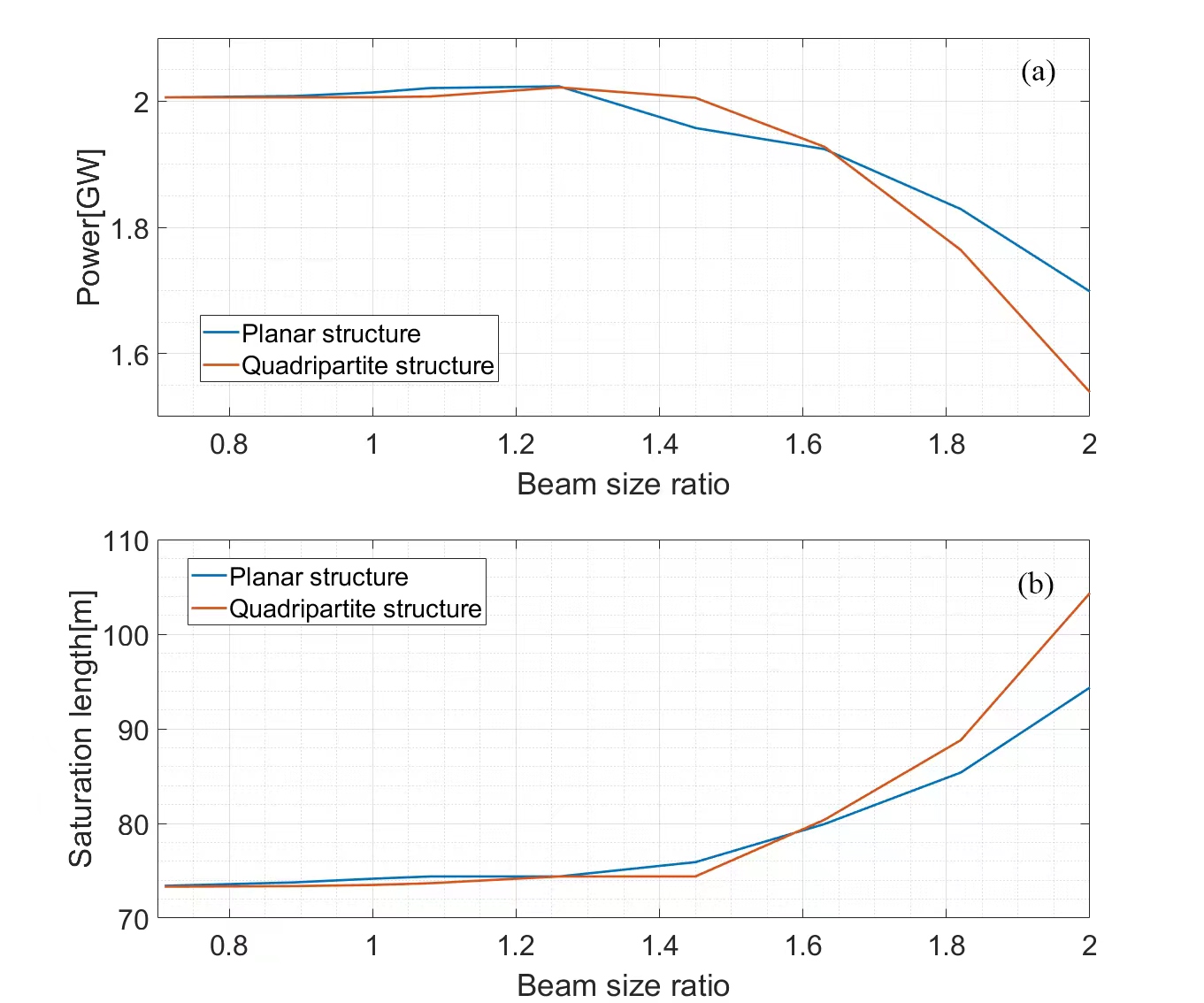}
	\caption{Saturation power (a) and saturation length (b) as a function of the beam size ratio.}
	\label{fig_lasingsaturation}
\end{figure}

The dependence of the saturation power and saturation length on the beam sizes is illustrated in Fig.~\ref{fig_lasingsaturation}. For both structures, emittance growth up to a beam size ratio of $\sim$1.5 results in only minor degradation of FEL performance. Beyond this threshold, the deterioration becomes more pronounced. At the largest beam size ratio of 2.0, the maximum output power decreases by 15.3\% for the planar structure and by 23.2\% for the quadripartite structure, relative to the 0.7 baseline case. Correspondingly, the saturation length increases by 28.5\% and 42.3\% for the two structures, respectively.

In conclusion, although the lasing performance is largely unaffected by higher-order and nonlinear components of the three-dimensional wakefields under the nominal beam conditions in Table~\ref{tab_beam}, the present study shows that these nonlinearities become non-negligible when the transverse beam size increases. Likewise, their influence could become significant for longer bunches. These findings highlight the necessity of including nonlinear wakefield effects in beam dynamics simulations, particularly for FEL applications where accurate knowledge of beam properties is critical.

\begin{figure*}[!htb]
	\includegraphics[width=\hsize]{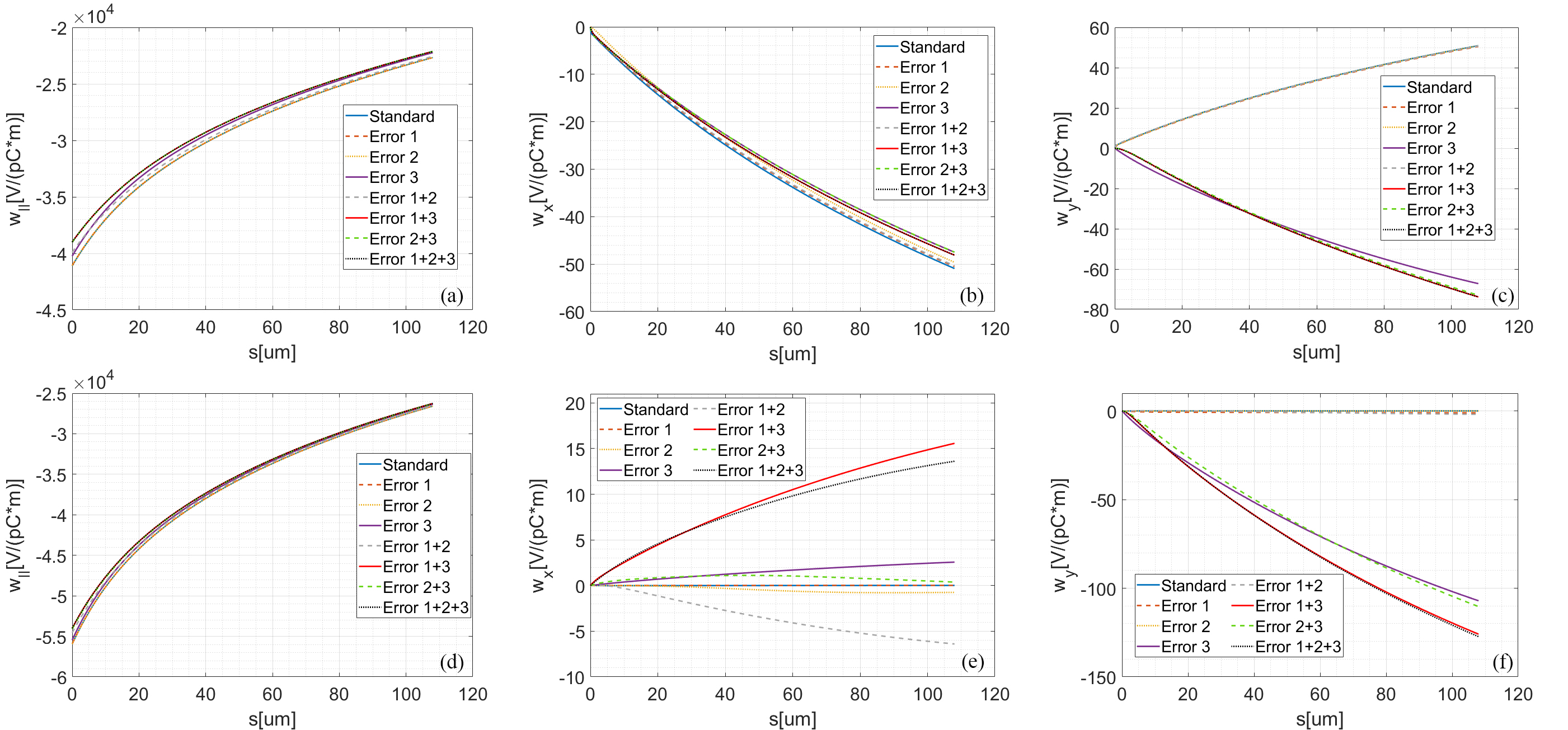}
	\caption{The on-axis wake functions of the planar (top) and the quadripartite (bottom) structures with various error types. The results of the standard case without errors are also provided for comparison. Left to right: longitudinal, horizontal, and vertical wake functions.}
	\label{fig_error_onaxis}
\end{figure*}

\section{Analysis of assembly errors}\label{sec.VIII}
Wakefield structures require accurate positioning of corrugated plates in order to obtain the designed wakefields. Therefore, it is essential to conduct analysis based on various types of assembly error to study the engineering feasibility, which is enabled by the developed particle-to-particle approach with three-dimensional wakefields.

Three typical types of shift errors that are likely to occur during assembly, as well as their combination, are considered in this study, as illustrated in Fig.~\ref{fig_crosssection}. The first type (denoted as Error~\#1) refers to the longitudinal misalignment of one plate along the beam traveling direction. The second type (denoted as Error~\#2) refers to the transverse misalignment of one plate perpendicular to the plate movement direction. The third type (denoted as Error~\#3) refers to the transverse misalignment of a single plate along the plate movement direction (outwards). Based on the engineering experience gained when assembling the prototype structure proposed in Ref.~\cite{Ji_LINAC2024}, the shift misalignment is set to 20~$\mu$m in all three types of error.

It should be noted that realistic assembly could present even more complicated types of error, such as plate tilt~\cite{Zhang_PRAB2015}, rotation, twist, and corrugation machining errors. Analysis of these types of error is beyond the scope of the current study and will be investigated in future work. 

Figure~\ref{fig_error_onaxis} illustrates the on-axis wakefield functions of structures with these error types obtained by \texttt{ECHO3D} simulation using an on-axis beam and deconvolution. While either Error~\#1 or Error~\#2 alone introduces negligible or modest impacts on the wakefields, Error~\#3 alone or with any combination of Error~\#1 or Error~\#2 remarkably changes the wakefields along the plate movement direction in both structures, reversing the original wakefield directions in the planar structure and inducing large transverse wakefields in the quadripartite structure.

\begin{figure}[!htb]
\includegraphics[width=\hsize]{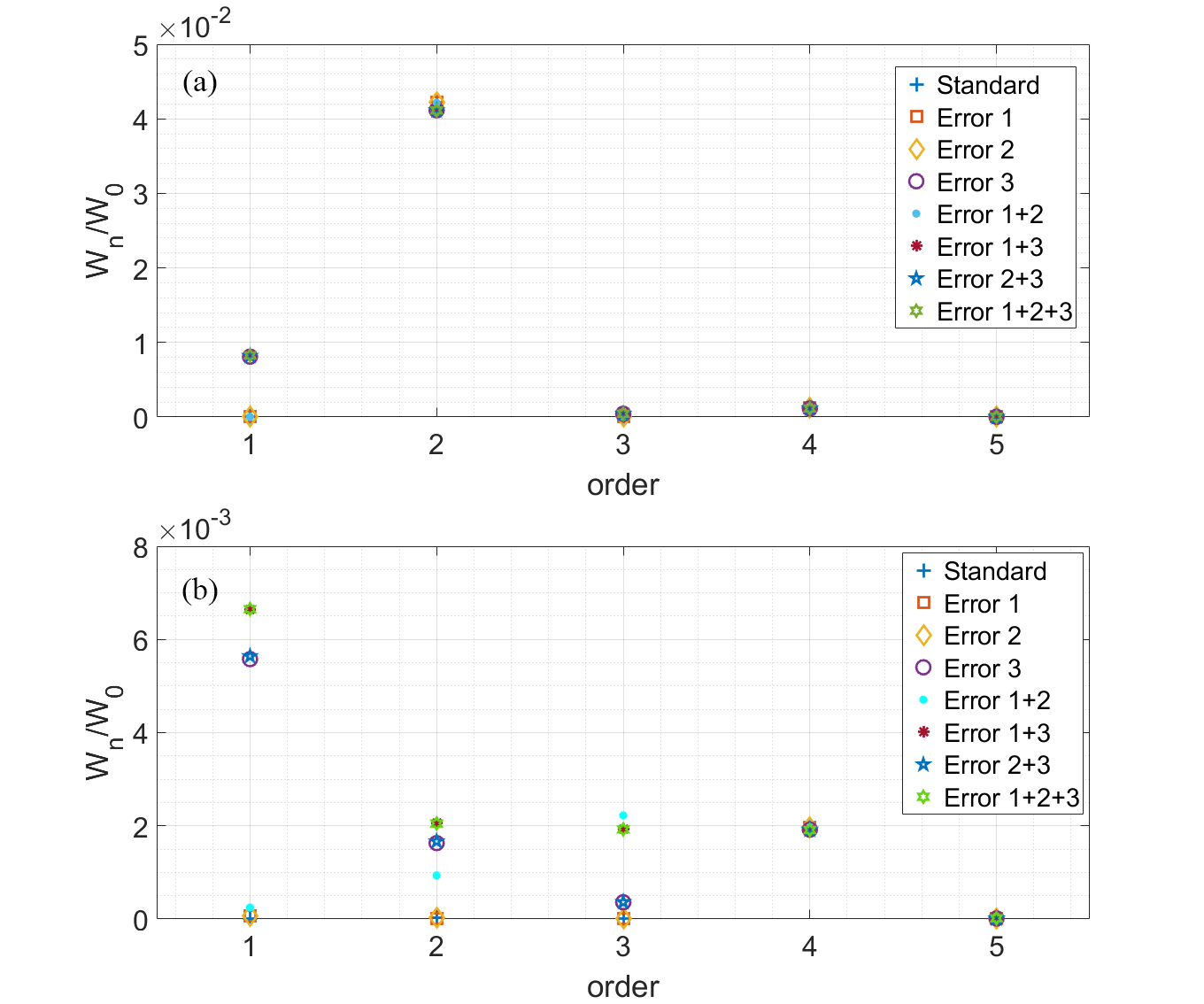}
\caption{The relative amplitude of the $n$th-order components with various error types for the planar (a) and the quadripartite (b) structures. The longitudinal wakefields with $r_0$=0, $r$=0.2~mm, and $s$=100~$\mu$m are expressed as the sum of a series of higher-order components.}
\label{fig_error_azimuthal}
\end{figure}

By analyzing the higher-order components via expressing the longitudinal wake functions azimuthally according to Eq.~\eqref{eq_azimuthal}, the large impact of Error \#3 is confirmed to be caused by dipole components, as illustrated in Fig.~\ref{fig_error_azimuthal}.

\begin{table*}[!htb]
	\centering
	\caption{Comparison of projected emittance growth under different structure configurations and error types.}
	\label{tab_emittance_errors}
	\begin{ruledtabular}
	\begin{tabular*}{16cm} {@{\extracolsep{\fill} } llcccc}
		Structures & Error type & $\epsilon_{x}$ (mm mrad) & $\epsilon_{y}$ (mm mrad) & $\Delta \epsilon_{x}(\%)$  & $\Delta \epsilon_{y}(\%)$ \\
		\hline
		\multirow{8}{*}{Planar structures (H+V)} & Standard & 0.618 & 0.519 & 31.54 & 25.41 \\
		& Error \#1 & 0.625 & 0.520 & 33.18  & 25.41 \\
		& Error \#2 & 0.624 & 0.519 & 32.93  & 25.29 \\
		& Error \#3 & 0.636 & 0.590 & 35.40  & 42.42 \\
		& Error \#1+2 & 0.622 & 0.520 &  32.55 & 25.58 \\
		& Error \#1+3 &  0.643 & 0.582 &  36.91 & 40.46  \\
		& Error \#2+3 &  0.642 & 0.579 &  36.72 & 39.81  \\
		& Error \#1+2+3 &  0.643 & 0.582 &  36.89 & 40.48  \\
		\multirow{8}{*}{Quadripartite structures} & Standard & 0.471 & 0.415 & 0.23 & 0.31 \\
		& Error \#1 & 0.471 & 0.415 & 0.23  & 0.34 \\
		& Error \#2 & 0.471 & 0.415 & 0.21  & 0.34 \\
		& Error \#3 & 0.476 & 0.584 & 1.30  & 40.94 \\
		& Error \#1+2 & 0.470 & 0.415 & 0.19 & 0.34 \\
		& Error \#1+3 & 0.476 & 0.599 & 1.41 & 44.69 \\
		& Error \#2+3 & 0.476 & 0.593 & 1.32 & 43.12 \\
		& Error \#1+2+3 & 0.476 & 0.600 & 1.38 & 45.00 \\
	\end{tabular*}
	\end{ruledtabular}
\end{table*}

The influence on the beam is then investigated by the particle-to-particle approach, where the results of projected emittance growth at the exit of the dechirper section are summarized in Table~\ref{tab_emittance_errors}. For both structures, the influence on beam emittance is negligible with Error \#1 and Error \#2, but significant when Error \#3 is introduced, consistent with the impacts on the on-axis wakefields as illustrated in Fig.~\ref{fig_error_onaxis}. 

\begin{figure}[!htb]
	\includegraphics[width=\hsize]{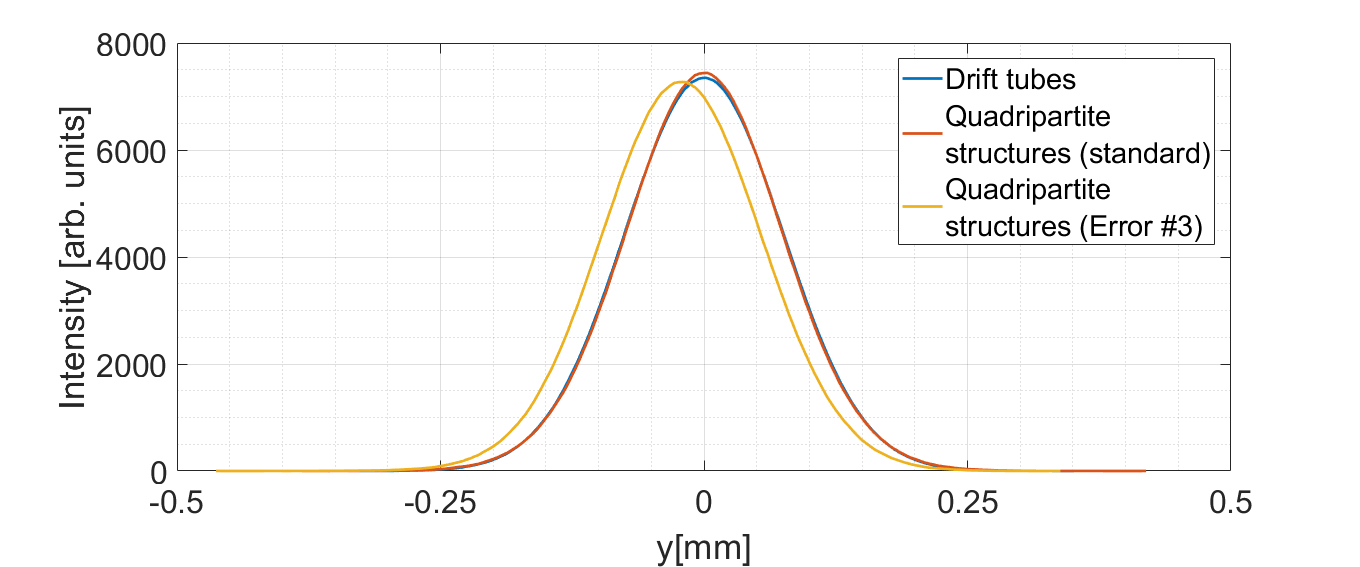}
	\caption{The transverse beam distribution in the vertical direction at 10~m downstream of the dechirper section.}
	\label{fig_offset_YAG}
\end{figure}

To make the quadripartite structure feasible for FEL facilities, it is critical to detect the presence of Error \#3 and control the misalignment. Since this type of assembly error leads to strong dipole wakefield, the beam offset after the structure detected by YAG profiles or beam position monitors can serve as an indicator of Error \#3. As illustrated in Fig.~\ref{fig_offset_YAG}, when the effective gap is closed from the wide-open position (equivalent to drifting) to the nominal setting, the beam center remains unchanged for the standard structure while shifts for the structure with Error \#3. Once detected, Error \#3 can be corrected by high-precision motors driving the corrugated plates with $\mu$m-scale accuracy.

Therefore, it can be concluded that the quadripartite structure is technically feasible for dechirper applications, where the most harmful type of assembly error can be detected and controlled during beam commissioning. 

\section{Preliminary mechanical design}\label{sec.IX}
\begin{figure}[!htb]
	\includegraphics[width=1\hsize]{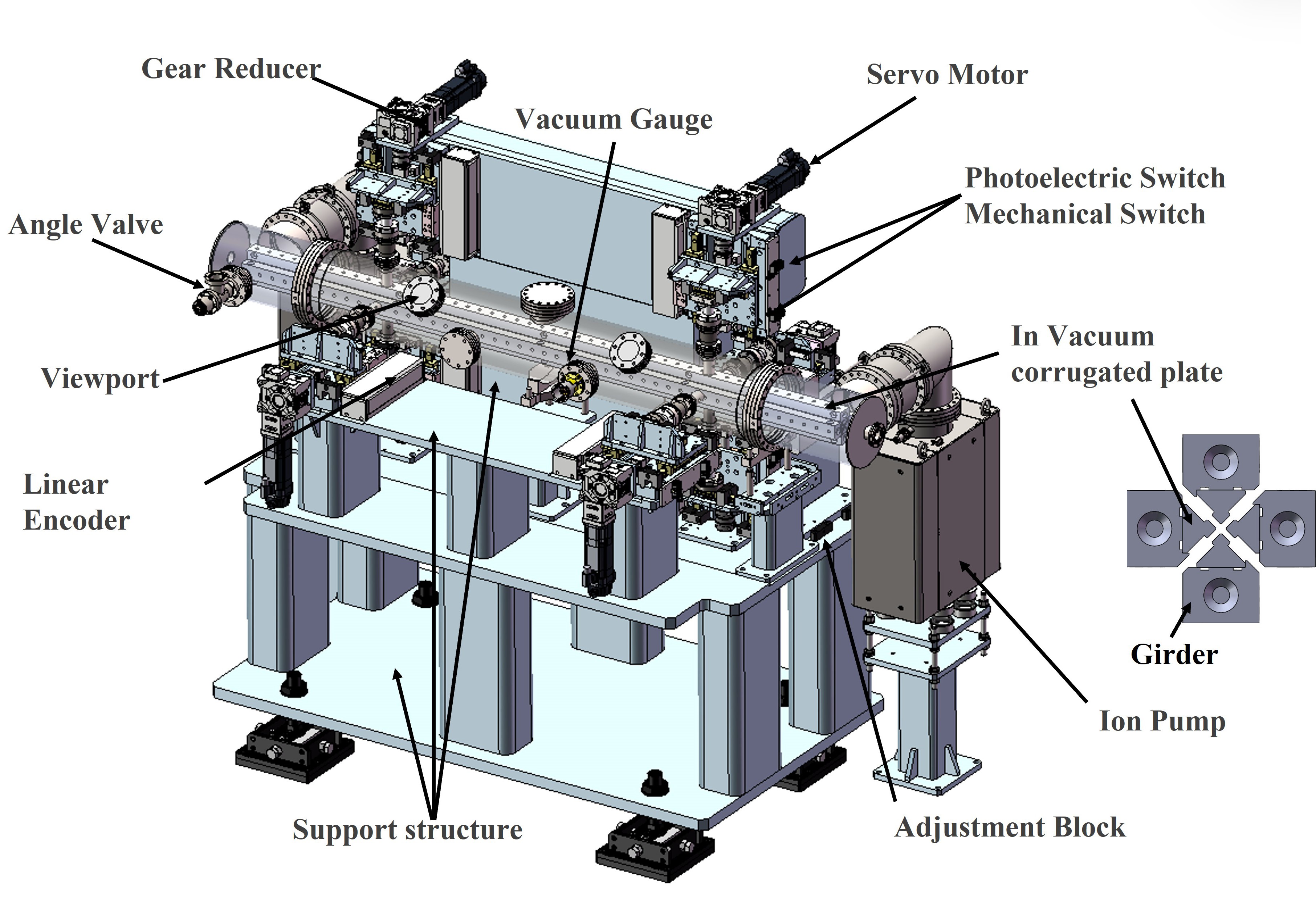}
	\caption{3D model of the quadripartite wakefield structure assembly. Inset: zoomed-in view of the girders holding the corrugated plates.}
	\label{fig_mechanical}
\end{figure}

The preliminary mechanical design of the quadripartite structure assembly is illustrated in Fig.~\ref{fig_mechanical}.

The assembly can be divided into four parts based on functional attributes, including the wakefield plates, the vacuum chamber, the transmission mechanism, and the supporting structure.

Each wakefield plate consists of corrugated segments secured onto a 2~m-long girder. The girder length is determined by the optimal wakefield length of 3.66~m plus a 10\% overhead. The corrugated plates are made of 6061 aluminum alloy to reduce the static deformation. To minimize cumulative machining errors and to meet tolerance requirements, the corrugated segments are individually machined to 25~cm long, with dimensional tolerances controlled within 20~$\mu$m. Venting slots are designed on mating surfaces between the girder and the corrugated segments to facilitate vacuum pumping.

The vacuum chamber provides a high-vacuum environment for the corrugated plates and its associated components. By using two ion pumps with 400~L/s pumping speed each, a high vacuum below 1x10$^{-6}$~Pa can be achieved according to \texttt{MOLFLOW+}~\cite{Kersevan_JVST2009} simulation.

The core of the entire assembly is the transmission mechanism with a high-precision control system, which realizes the linear reciprocating movement of the wakefield plates, thereby adjusting the effective gap. The transmission system consists of a two-dimensional motion platform assembled with two sets of linear guides. The primary platform is motor-driven to achieve linear motion of the girder. The secondary transmission is manually controlled by adjusting the jacking screws on both side blocks, which drives the wakefield plates to move perpendicular to the primary drive direction and enables free extension during vacuum baking. Additionally, a multi-degree-of-freedom adjustment mechanism is designed between the slider and the hanger rod flange, allowing for posture adjustment of the corrugated plates to ensure assembly accuracy. 

The control system integrates a high-resolution linear encoder for position-loop feedback with a servo motor's velocity-loop control, and enables full closed-loop operation with both precision positioning and dynamic response. The transmission mechanism is equipped with four-layer motion protection, including soft limits implemented in the control system, photoelectric limit switches, mechanical limit switches, and hard limits by blocks attached to the primary linear guides.

The support structure comprises upper and lower brackets, securely connected via bolts combined with dowel pins. The lower bracket features leveling feet enabling positional adjustments ($\pm$20 mm) along transverse and longitudinal axes. The upper bracket incorporates push-pull adjustment blocks for fine positioning. This support system carries the wakefield plates, the vacuum chamber, and the transmission mechanism, thereby providing reliable assurance of stable operation.

\section{Conclusion}\label{sec.XI}
In this manuscript, a quadripartite structure is proposed as a dechirper in FELs. The design has the flexibility of tuning the longitudinal wakefield for various beam conditions by changing the gaps between the corrugation plates. Compared with the conventional planar design, the structure offers enhanced longitudinal wakefield and fully suppressed quadrupole wakefields when the beam traverses on axis. Therefore, the structure is suitable as dechirper by providing shortened length and preserved emittance.

To further evaluate the structure performance in the presence of higher-order components and nonlinear effects in realistic wakefields, a particle-to-particle method is developed with three-dimensional wakefields obtained by \texttt{ECHO3D} simulation and deconvolution. The simulation results confirm that the impact of these nonlinear effects is negligible for the quadripartite design under nominal beam conditions. However, the nonlinear effects can lead to remarkable degradation of beam slice emittance and lasing performance when using larger beam sizes, which highlights the necessity of including three-dimensional wakefields in beam dynamics simulations.

The methodology also enables analysis of structure assembly errors where the positioning error along the plate movement direction is found to introduce large dipole wakefields and lead to significant emittance growth. However, such error can be detected and controlled by beam-based alignment, which proves the engineering feasibility of the proposed structure.

In addition to its use as a dechirper, the quadripartite structure can also be applied for beam manipulation and diagnostics in FELs, ultrafast electron diffraction/imaging setups~\cite{Fu_PRL2015,Lu_PRAB2016}, compact light sources~\cite{Dong_NIMA2021}, and test facilities~\cite{Antipov_PRL2013,Antipov_PRL2014,Nie_NIMA2016,Pacey_NIMA2018,Ha_RMP2022}.

Recently, a prototype structure featuring 1~m-long corrugation plates has been developed for beam tests to benchmark the simulation results and validate engineering aspects~\cite{Ji_LINAC2024}. The structure has been fully assembled and the experiment will soon be conducted at the undulator beamline of Dalian Coherent Light Source~\cite{Sun_NIMA2024,Li_PRAB2025}.

More types of structure errors and issues related to high-repetition-rate operation (heat and radiation control~\cite{Bane_PRAB2017}, beam loss mitigation~\cite{Guo_NIMA2022}, etc.) will be investigated for applications of the quadripartite structure in S$^3$FEL.

\begin{acknowledgments}
The authors would like to thank strong supports from Dalian Coherent Light Source, fruitful discussions with Dr. Chao Feng and Dr. Zhen Wang on wakefield structure study, and invaluable help from Dr. Igor Zagorodnov on \texttt{ECHO3D} simulation. This work is supported by the Talent Program of Guangdong Province (No.2021QN02G685).
\end{acknowledgments}

\appendix
\section{Benchmarking of ECHO3D against CST}\label{ECHO}
Two input files are required for the \texttt{ECHO3D} simulation: one defines the structure geometry, and the other specifies the beam parameters, mesh settings, boundary conditions, and solver configurations.

Since the wakefield structures investigated in this study are periodic (either ideal or with shift alignment errors), only the vacuum region of a single period is modeled in CAD software and then replicated within the \texttt{ECHO3D} mesher. This approach significantly reduces meshing time and prevents potential code failures. The boundary conditions are set to electric, while the “simple conformal” meshing scheme and the “implicit” solver are employed. Further details regarding the code configuration can be found in the user manual~\cite{ECHO3DManual}.

In this study, \texttt{ECHO3D} is benchmarked with the wakefield solver in \texttt{CST}. Due to the extremely high computational cost of \texttt{CST}, it is impractical to apply the converged mesh sizes from Sec.~\ref{sec.III.A} and the nominal beam parameters listed in Table~\ref{tab_beam}. Instead, coarser mesh sizes of ($dx, dy, dz$) = (20~$\mu$m, 20~$\mu$m, 10~$\mu$m), a shorter structure length of 0.1~m, and a longer bunch with Gaussian temporal distribution and RMS length of 200~$\mu$m are adopted in both codes to compare the quadripartite structure.

\begin{figure}[!htb]
	\centering
	\includegraphics[width=\hsize]{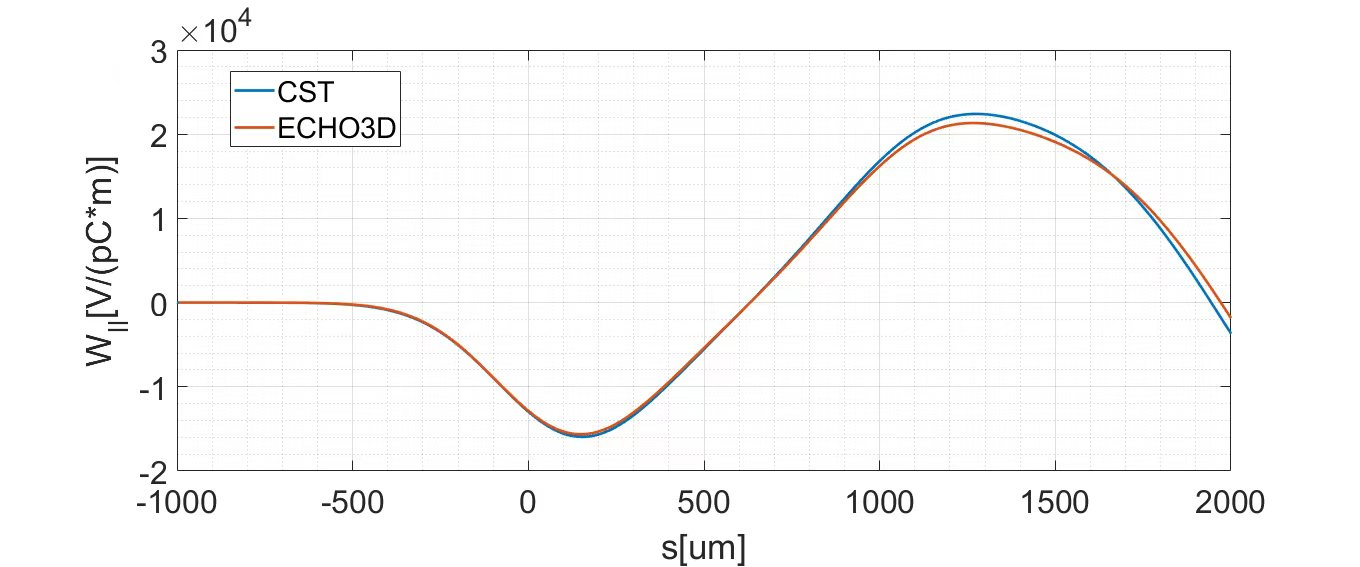}
	\caption{Comparison of the longitudinal wake potential of the ideal quadripartite structure.}
	\label{fig_comparison_ideal}
\end{figure}

\begin{figure}[!htb]
	\centering
	\includegraphics[width=\hsize]{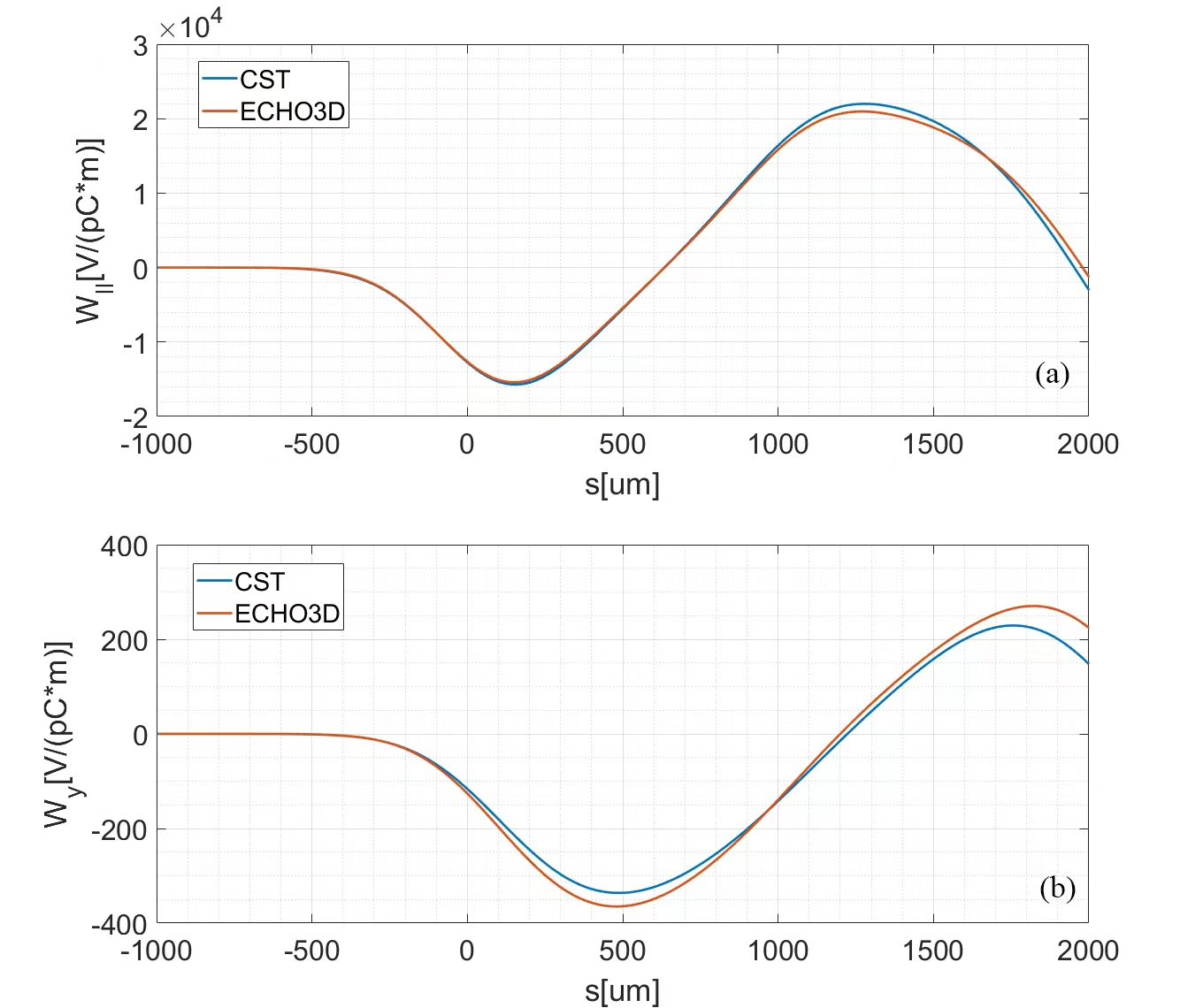} \caption{Comparison of the longitudinal (a) and vertical (b) wake potentials of the quadripartite structure with Error \#3.}
	\label{fig_comparison_Error3}
\end{figure}

The results of the ideal structure and the one with Error \#3 are illustrated in Fig.~\ref{fig_comparison_ideal} and Fig.~\ref{fig_comparison_Error3}, respectively. The transverse wake potential obtained from \texttt{CST} is directly provided by the software, whereas that from \texttt{ECHO3D} is calculated using Eq.~\eqref{eq_PW}. The good agreement confirms the validity of both the \texttt{ECHO3D} simulation and the post-processing procedure.
	
\section{Impact of corrugation width on wakefields of the planar structure}\label{structruewidth}
\begin{figure}[!htb]
	\centering
	\includegraphics[width=\hsize]{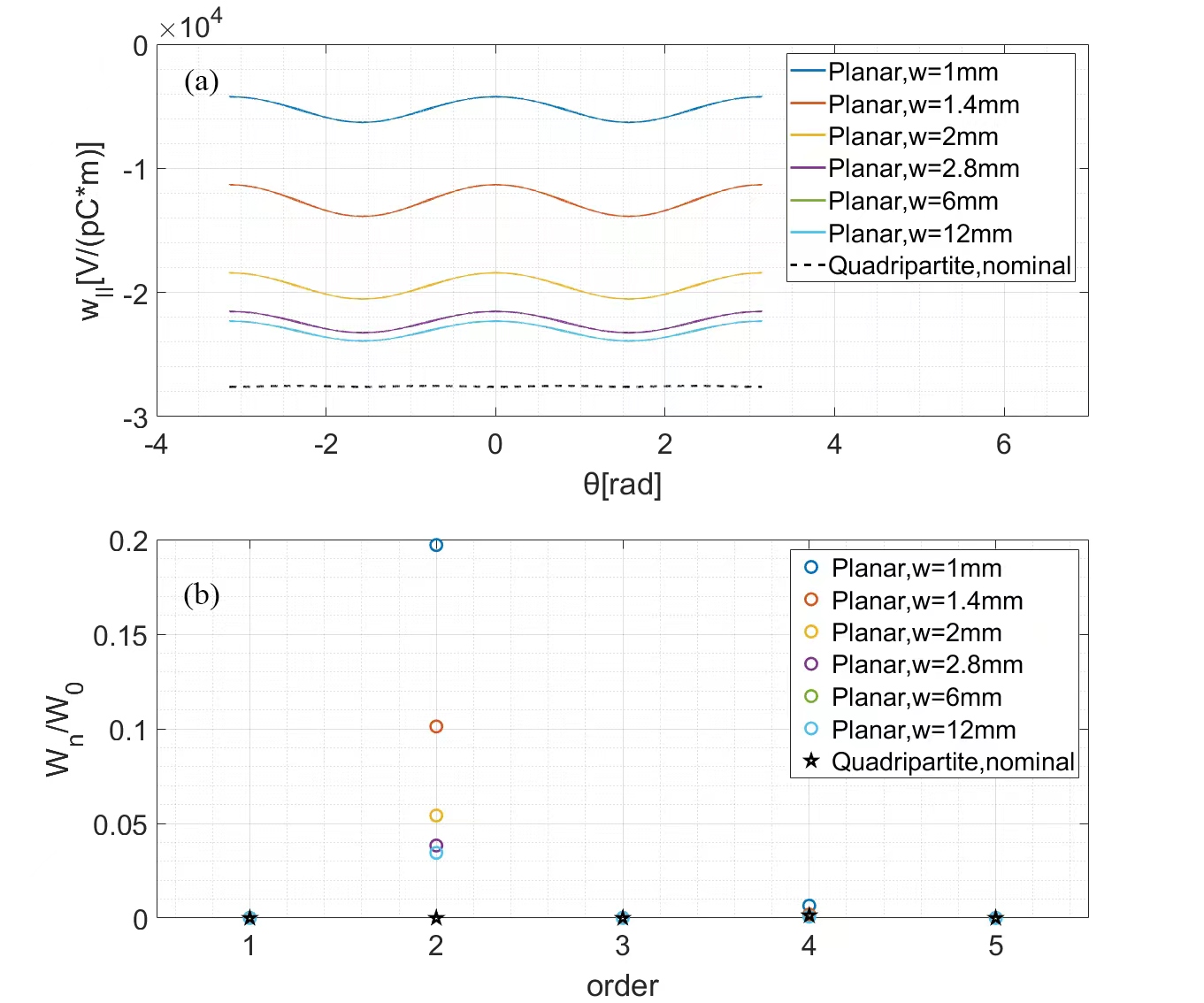}
	\caption{(a) The longitudinal wake function along the azimuthal direction. (b) The amplitude of the higher-order components normalized by the $0$th-order one.}
	\label{fig_azimuthal_width}
\end{figure}

The corrugation width of the planar structure is varied to investigate the impact on wakefields by \texttt{ECHO3D} simulations. Following the post-processing procedure in Sec.~\ref{sec.V}, the longitudinal wakefields at $r_0$=0, $r$=0.2~mm, $s$=100~$\mu$m are illustrated in Fig.~\ref{fig_azimuthal_width}. The longitudinal wakefield nearly remains constant for $w$ is larger than 3~mm, and rapidly decreases with smaller $w$. When $w$ is equal to the nominal effective gap of 1.4~mm, the longitudinal wakefield of the planar structure is only about half of the quadripartite structure.

\begin{figure}[!htb]
	\centering
	\includegraphics[width=\hsize]{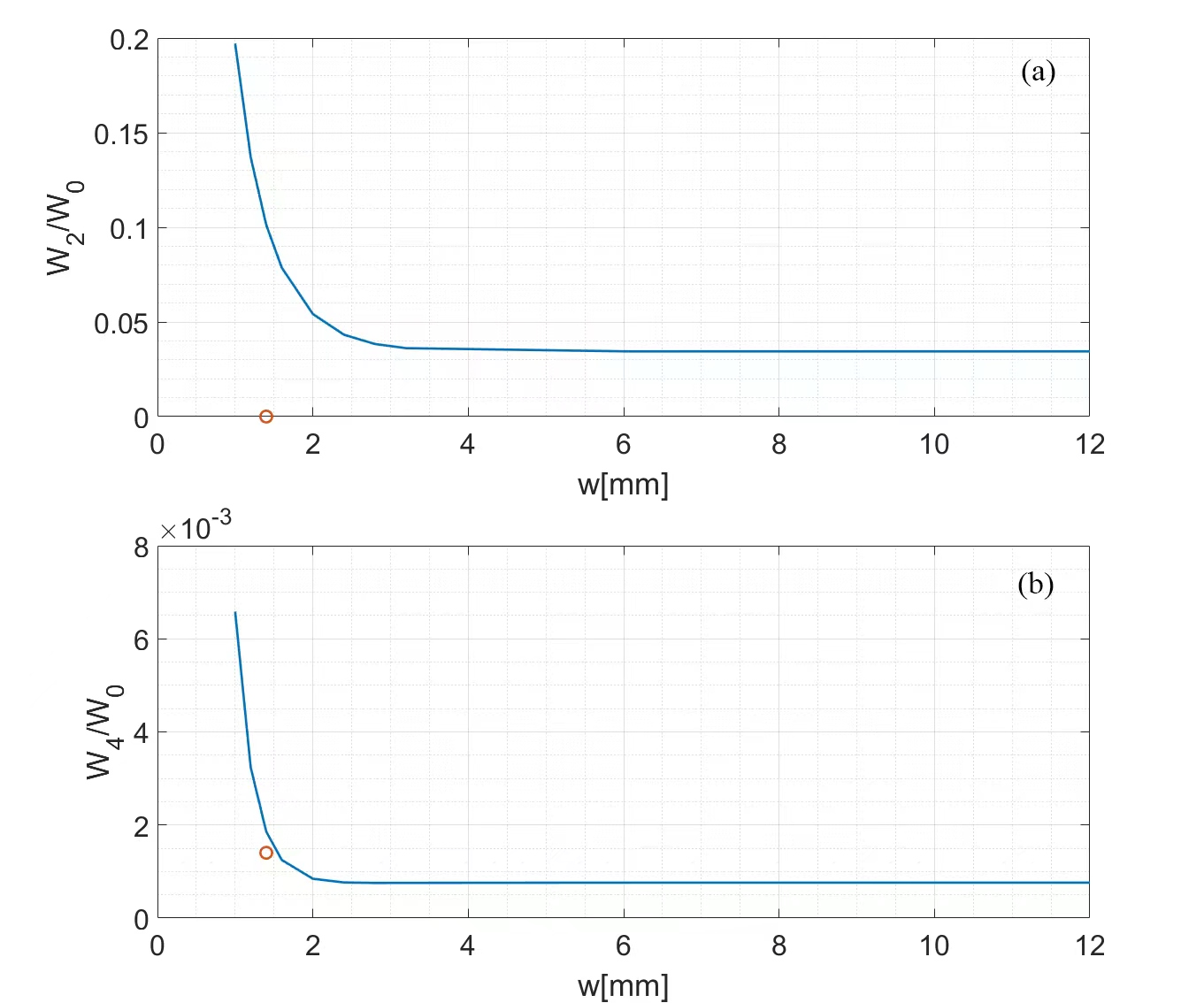}
	\caption{The normalized quadrupole (a) and octupole (b) components of planar structure as a function of corrugation width. The values of the quadripartite structure at the nominal effective gap is marked by the red circles for reference.}
	\label{figure_higherorder_width}
\end{figure}

The dependence of the higher-order components on $w$ is illustrated in Fig.~\ref{figure_higherorder_width}. Stronger quadrupole and octupole components can be observed with smaller $w$. When $w$ is equal to the nominal effective gap of 1.4~mm, the octupole component of the planar structure is comparable to that of the quadripartite structure.

\section{Validation of the first assumption in the particle-to-particle tracking method}\label{app}
Since it is impractical to run \texttt{ECHO3D} simulation for every particle with different offsets in the beam, the entire beam is divided into macro-particles by transverse positions so that Eq.~\eqref{eq_p2p_approximation} can be simplified as
\begin{equation}
	\label{eq_p2p_approximation_divide}
	\left\{
	\begin{aligned}
		\frac{1}{N}\sum_{j} N_j w_{||}\left(x_{j}, x, y_{j}, y, s\right) \approx w_{||}\left(\overline{x}, x, \overline{y}, y, s\right)\\
		\frac{1}{N}\sum_{j} N_j w_{x}\left(x_{j}, x, y_{j}, y, s\right) \approx w_{x}\left(\overline{x}, x, \overline{y}, y, s\right)\\
		\frac{1}{N}\sum_{j} N_j w_{y}\left(x_{j}, x, y_{j}, y, s\right) \approx w_{y}\left(\overline{x}, x, \overline{y}, y, s\right)
	\end{aligned}
	\right.,
\end{equation}
where $N_j$ and ($x_j, y_j$) denote the number of particles and the average offset of the $j$-th macro-particle.

In addition, wake potentials instead of wake functions are used to further reduce the computation cost of deconvolution. Therefore, Eq.~\eqref{eq_p2p_approximation_divide} is equivalent to
\begin{equation}
	\label{eq_p2p_approximation_divide_potential}
	\left\{
	\begin{aligned}
		\frac{1}{N}\sum_{j} N_j W_{||}\left(x_{j}, x, y_{j}, y, s\right) \approx W_{||}\left(\overline{x}, x, \overline{y}, y, s\right)\\
		\frac{1}{N}\sum_{j} N_j W_{x}\left(x_{j}, x, y_{j}, y, s\right) \approx W_{x}\left(\overline{x}, x, \overline{y}, y, s\right)\\
		\frac{1}{N}\sum_{j} N_j W_{y}\left(x_{j}, x, y_{j}, y, s\right) \approx W_{y}\left(\overline{x}, x, \overline{y}, y, s\right)
	\end{aligned}
	\right..
\end{equation}

\begin{figure}[!htb]
\includegraphics[width=0.8 \hsize]{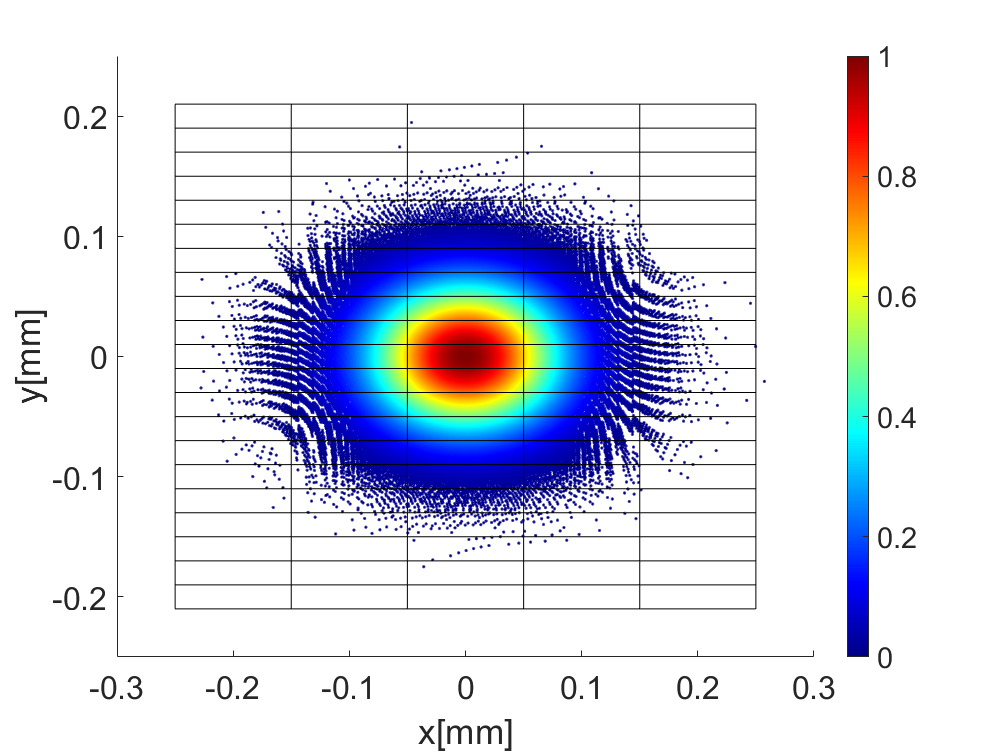}
\caption{The transverse profile of the beam used in the appendix. The grids denote the boundaries of macro-particles.}
\label{fig_transverse_profile}
\end{figure}

In this study, the beam at the entrance of the dechirper section is adopted, where the transverse distribution is preserved and the longitudinal profile is re-distributed following Gaussian shape with RMS bunch length of 15~$\mu$m. Figure~\ref{fig_transverse_profile} illustrates the transverse profile of the beam, together with the grids to determine macro-particles. 

\begin{figure*}[!htb]
	\includegraphics[width=\hsize]{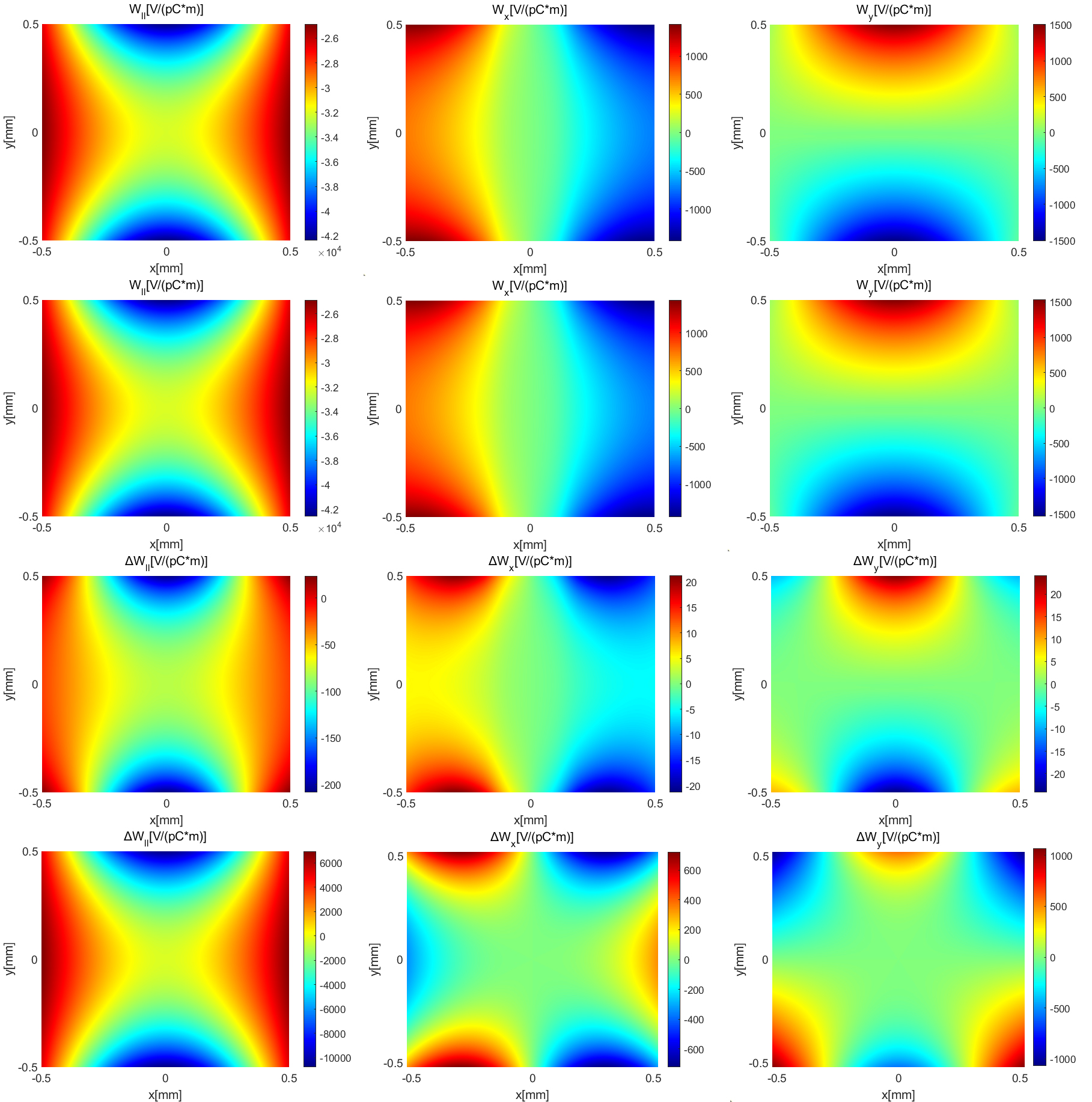}
	\caption{Simulation results of the planar structure with $g=$1.4~mm and the original beam size. Left to right: longitudinal, horizontal, and vertical wake functions.}
	\label{fig_app_3D_A}
\end{figure*}

\begin{figure*}[!htb]
	\includegraphics[width=\hsize]{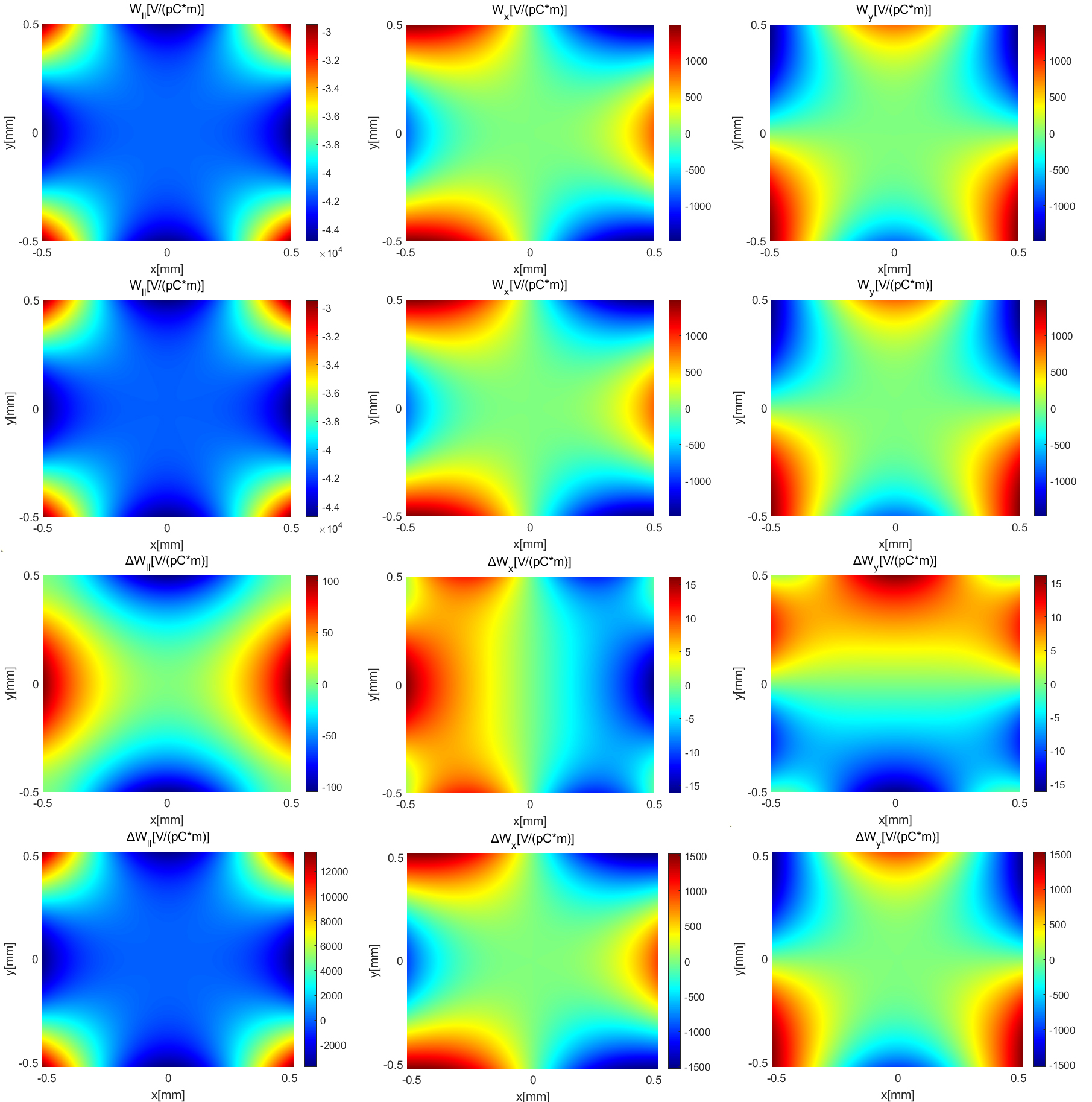}
	\caption{Simulation results of the quadripartite structure with $g_x=g_y=$1.4~mm and the original beam size. Left to right: longitudinal, horizontal, and vertical wake functions.}
	\label{fig_app_3D_B}
\end{figure*}

The \texttt{ECHO3D} simulation results of the planar and the quadripartite structure with nominal effective gap are respectively illustrated in Fig.~\ref{fig_app_3D_A} and Fig.~\ref{fig_app_3D_B}, where the slice wakefields at $s$=100~$\mu$m are presented. The first and the second row denotes the left-hand and the right-hand side of Eq.~\eqref{eq_p2p_approximation_divide_potential}, while the third row denotes their differences. The fourth row denotes the differences between the ideal wakefields and the realistic wakefields in Fig.\ref{fig_3D_ideal} and Fig.~\ref{fig_3D_realistic}, convoluted by the Gaussian temporal distribution. It can be observed that top two rows closely match each other, and their differences are significantly smaller than that between one-dimensional and three-dimensional wakefields. Therefore, the approximation in Eq.~\eqref{eq_p2p_approximation} is justified.

\begin{figure*}[!htb]
	\includegraphics[width=\hsize]{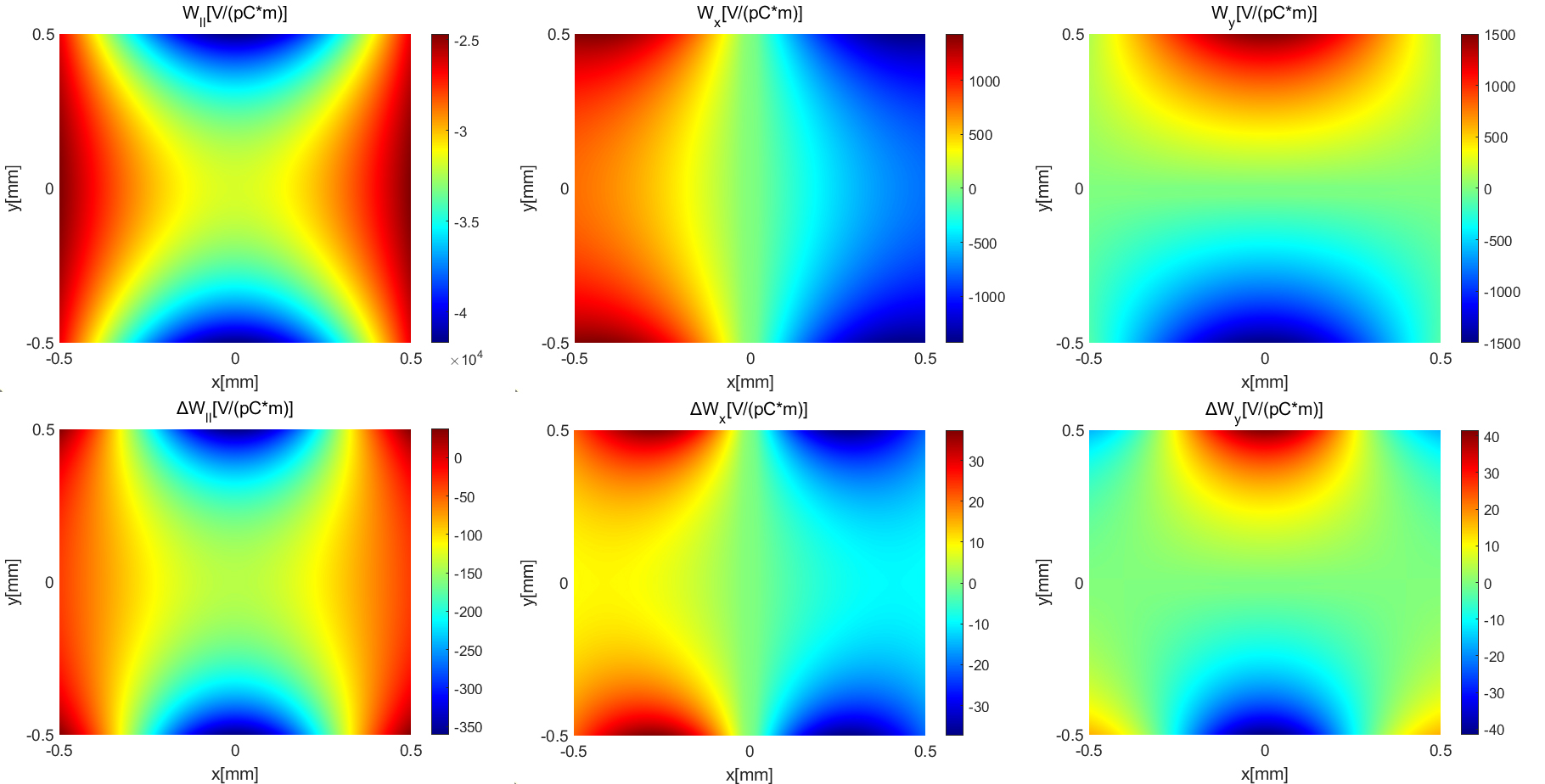}
	\caption{Simulation results of the planar structure with $g=$1.4~mm and the enlarged beam size. Left to right: longitudinal, horizontal, and vertical wake functions.}
	\label{fig_app_3D_A_2}
\end{figure*}

\begin{figure*}[!htb]
	\includegraphics[width=\hsize]{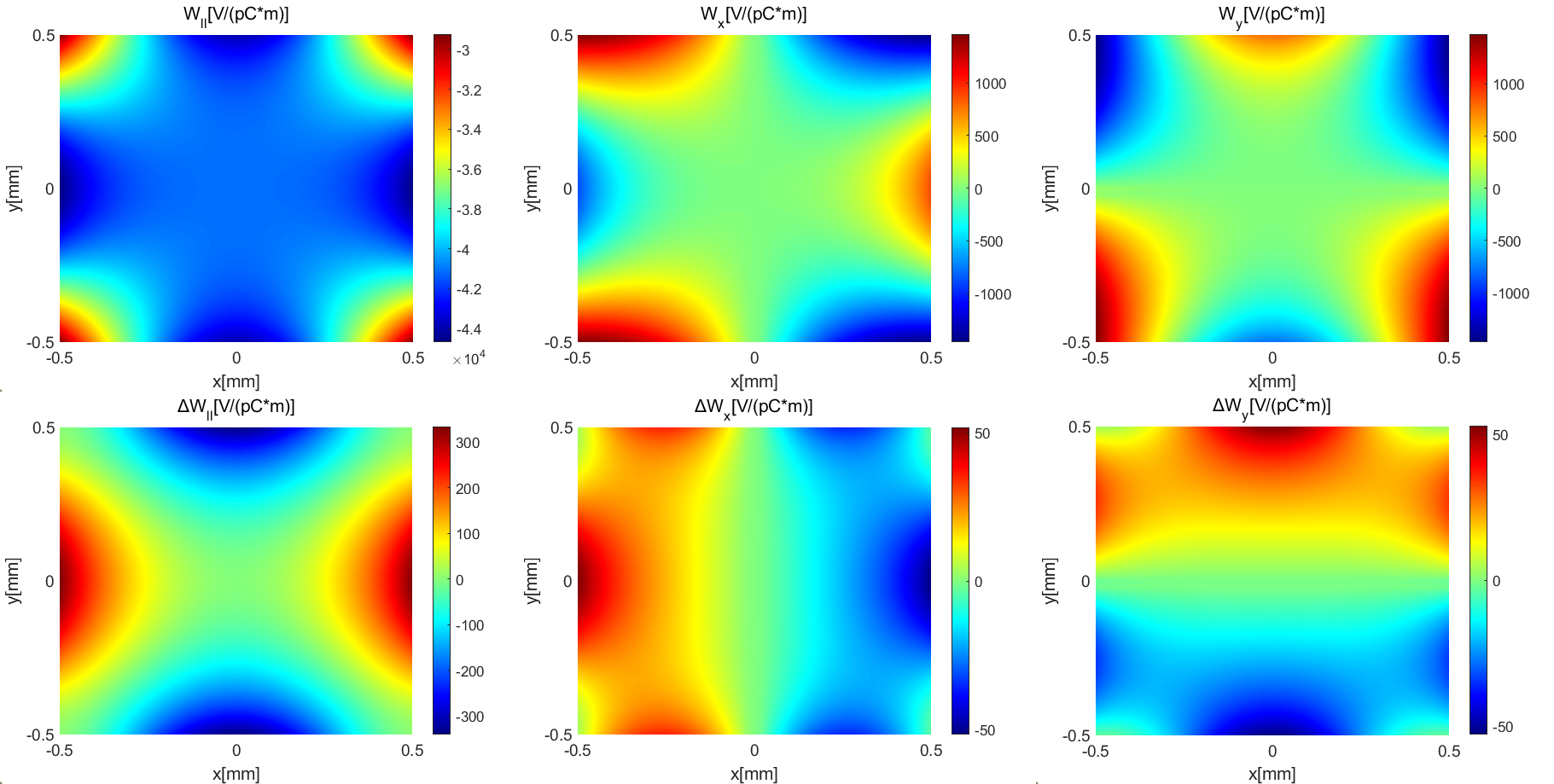}
	\caption{Simulation results of the quadripartite structure with $g_x=g_y=$1.4~mm and the enlarged beam size. Left to right: longitudinal, horizontal, and vertical wake functions.}
	\label{fig_app_3D_B_2}
\end{figure*}

To further examine the range of validity of the approximation, the beam size is increased by scaling the transverse positions of particles in the original distribution by a factor of two. The updated \texttt{ECHO3D} simulation results of the planar and the quadripartite structure are respectively illustrated in Fig.~\ref{fig_app_3D_A_2} and Fig.~\ref{fig_app_3D_B_2}, where the slice wakefields at $s$=100~$\mu$m are presented. The first row displays the averaged wake potentials of the updated macro-particles, and the second row shows their deviations from the wake potential of the on-axis particle. Although these deviations are several times larger than those for the original beam size, they remain significantly smaller than the discrepancies between the one-dimensional and three-dimensional wakefields, confirming the validity of the approximation in Eq.~\eqref{eq_p2p_approximation} for the enlarged beam.

\makeatletter
\let\auto@bib@innerbib\@empty
\makeatother

\end{document}